\documentclass[
reprint,
superscriptaddress,
nofootinbib,
longbibliography
amsmath,
amssymb,
aps,
prb,
]{revtex4-2}
\usepackage{xcolor}
\definecolor{darkblue}{rgb}{0,0,.6}
\usepackage[colorlinks=true, citecolor=darkblue, linkcolor=darkblue, menucolor=darkblue, urlcolor=darkblue]{hyperref}
\usepackage{float}

\usepackage{graphicx}% Include figure files
\usepackage{physics}
\usepackage{dcolumn}% Align table columns on decimal point
\usepackage{bm}% bold math
\usepackage{multirow}
\usepackage{amsmath}
\usepackage{array}
\usepackage[capitalize]{cleveref}

\usepackage[caption=false]{subfig}
\newcommand{\phantomsubfloat}[1]{
    {% apply caption setup only temporarily
        \captionsetup[subfigure]{labelformat=empty}
        \subfloat[][]{#1}
    }%
}

\newcommand{\BYRO}{$\mathrm{Ba_2YReO_6}$}
\newcommand{\SCRO}{$\mathrm{Sr_2CrReO_6}$}
\newcommand{\CFRO}{$\mathrm{Ca_2FeReO_6}$}
\newcommand{\BFRO}{$\mathrm{Ba_2FeReO_6}$}

\newcommand{\jes}{$j_{eff}=1/2$}
\newcommand{\jgs}{$j_{eff}=3/2$}
\newcommand{\Jtwo}{$J_{eff}=2$}
\newcommand{\jeff}{$j_{eff}$}
\newcommand{\leff}{$l_{eff}$}
\newcommand{\ltg}{$l_{eff}=1$}
\newcommand{\Ltg}{$L_{eff}=1$}
\newcommand{\tg}{$t_{2g}$}
\newcommand{\eg}{$e_g$}
\newcommand{\secBYRO}{\texorpdfstring{$\mathbf{Ba_2YReO_6}$}{TEXT}}
\newcommand{\secSCRO}{\texorpdfstring{$\mathbf{Sr_2CrReO_6}$}{TEXT}}
\newcommand{\secVdII}{\texorpdfstring{$\bm{5d^2}$}{TEXT}}
\newcommand{\secLII}{\texorpdfstring{$\bm{L_2}$}{TEXT}}
\newcommand{\secLIII}{\texorpdfstring{$\bm{L_3}$}{TEXT}}

\newcolumntype{C}[1]{>{\centering\let\newline\\\arraybackslash\hspace{0pt}}m{#1}}

\begin{document}

\preprint{APS/123-QED}

\title{Resonant inelastic X-ray scattering investigation of Hund's and spin-orbit coupling in \secVdII{} double perovskites}% Force line breaks with \\

\author{Felix I. Frontini}
 \email{ffrontini@anl.gov}
 \thanks{Present address: Advanced Photon Source, Argonne National Laboratory, 9700 S. Cass Avenue, Lemont, IL 60439}
 \affiliation{Physics Department, University of Toronto, 60 St. George Street, Toronto, ON, M5S 1A7}
\author{Christopher J. S. Heath}
 \affiliation{Physics Department, University of Toronto, 60 St. George Street, Toronto, ON, M5S 1A7}
\author{Bo Yuan}
 \thanks{Present address: Department of Physics and Astronomy, McMaster University, Hamilton, Ontario L8S 4M1, Canada}
 \affiliation{Physics Department, University of Toronto, 60 St. George Street, Toronto, ON, M5S 1A7}
\author{Corey M. Thompson}
    \thanks{Present address: Department of Chemistry, Purdue University, 560 Oval Drive, West Lafayette, Indiana 47907-2084, USA}
    \affiliation{Department of Chemistry and Chemical Biology, McMaster University, Hamilton, Ontario, Canada L8S 4L8}
\author{John Greedan}
    \affiliation{Department of Chemistry and Chemical Biology, McMaster University, Hamilton, Ontario, Canada L8S 4L8}
    \affiliation{Brockhouse Institute for Materials Research, McMaster University, Hamilton, Ontario, Canada L8S 4L8}
\author{Adam J. Hauser}
    \affiliation{Department of Physics and Astronomy, The University of Alabama, Tuscaloosa, Alabama 35487, USA}
\author{F. Y. Yang}
    \affiliation{Department of Physics, The Ohio State University, Columbus, Ohio 43210, USA}
\author{Mark P. M. Dean}
    \affiliation{Department of Condensed Matter Physics and Materials Science, Brookhaven National Laboratory, Upton, New York 11973, USA}
\author{Mary H. Upton}
 \affiliation{Advanced Photon Source, Argonne National Laboratory, 9700 S. Cass Avenue, Lemont, IL 60439}
\author{Diego M. Casa}
 \affiliation{Advanced Photon Source, Argonne National Laboratory, 9700 S. Cass Avenue, Lemont, IL 60439}
\author{Young-June Kim}%
 \email{youngjune.kim@utoronto.ca}
 \affiliation{Physics Department, University of Toronto, 60 St. George Street, Toronto, ON, M5S 1A7}%

\date{\today}% It is always \today, today,

\begin{abstract}
B site ordered $5d^2$ double perovskites ($\mathrm{A_2BB'O_6,\ B'}=5d^2)$ display a remarkable range of physical properties upon variation of the chosen B and $\mathrm{B'}$ site ions.
This sensitivity to chemical substitution reflects the delicate balance and profound impact of strong electronic correlation and spin-orbit coupling in such systems.
We present rhenium $L_2$ and $L_3$ resonant inelastic X-ray scattering (RIXS) measurements of two such physically dissimilar materials, Mott-insultating \BYRO{} and semiconducting \SCRO{}.
Despite these differences, our RIXS results reveal similar energy scales of Hund's ($J_H$) and spin-orbit coupling ($\zeta$) in the two materials, with both systems firmly in the intermediate Hund's coupling regime ($\order{J_H/\zeta}\sim 1$).
However, there are clear differences in their RIXS spectra. The conductive character of \SCRO{} broadens and obfuscates the atomic transitions within an electron-hole continuum, while the insulating character of \BYRO{} results in sharp atomic excitations.
This contrast in their RIXS spectra despite their similar energy scales reflects a difference in the itinerancy-promoting hopping integral and illustrates the impact of the local crystal environment in double perovskites.
Finally, $L_2$ and $L_3$ edge analyses of the atomic excitations in \BYRO{} reveal that the ordering of the low lying excited states is inverted compared to previous reports, such that the appropriate energy scales of Hund's and spin-orbit coupling are significantly modified.
We present exact diagonalization calculations of the RIXS spectra at both edges which show good agreement with our results for new energy scales of $\zeta=0.290(5)$ eV and $J_H=0.38(2)$ eV ($J_H/\zeta=1.30(5)$).
\begin{description}

\item[Usage]
Secondary publications and information retrieval purposes.
\end{description}
\end{abstract}

\maketitle

\section{\label{sec:intro}Introduction}
\begin{figure}[h]
\includegraphics[width=\columnwidth]{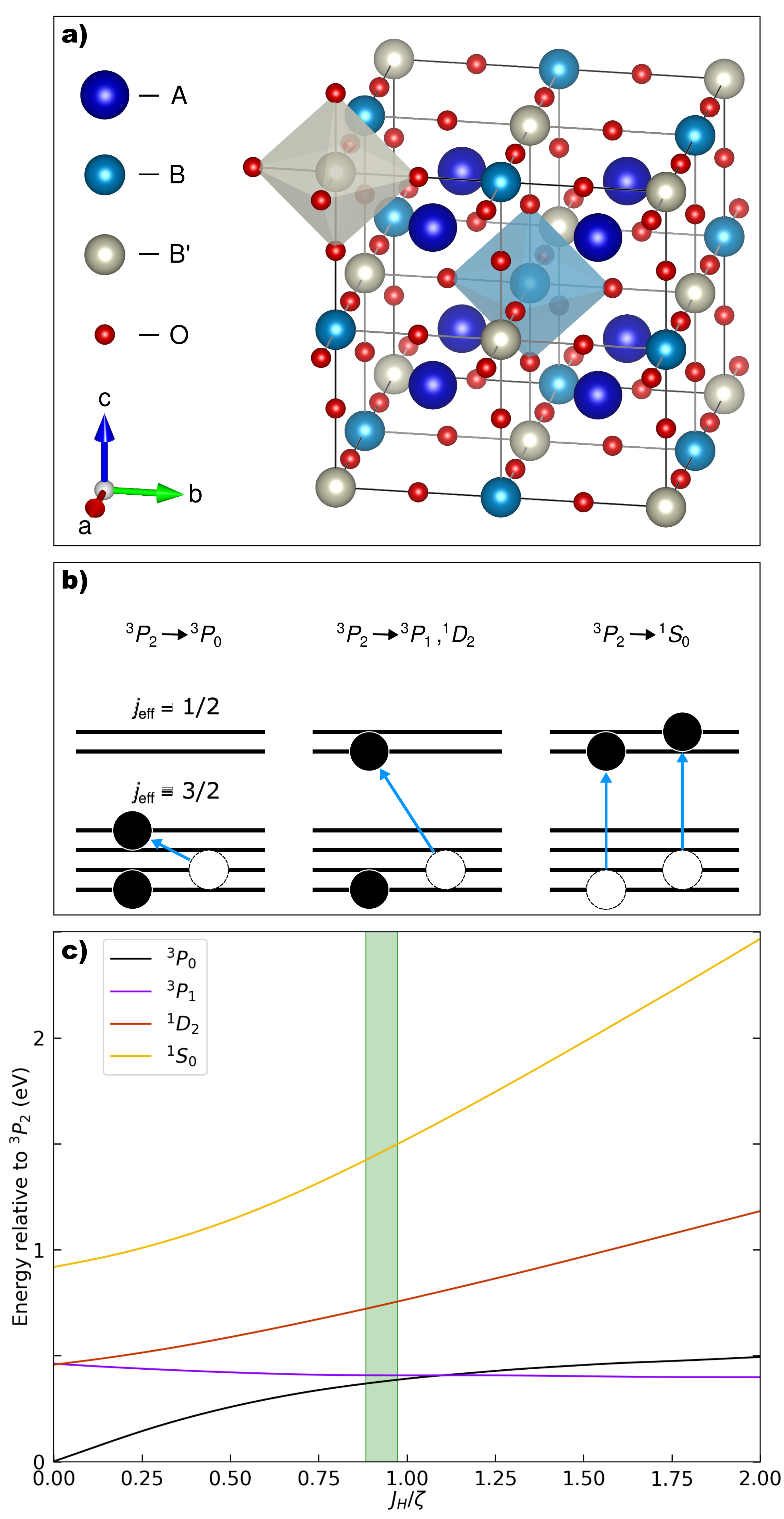}
\phantomsubfloat{\label{fig:fig1a}}
\phantomsubfloat{\label{fig:fig1b}}
\phantomsubfloat{\label{fig:fig1c}}
\vspace{-2\baselineskip}
\caption{a) Ideal cubic crystal structure of the B site ordered double perovskites. b) Schematic of $5d^2$ excitations from the single-particle perspective. c) $5d^2$ excited state energy level diagram as a function of $J_H$ for $\zeta =0.29$ eV, calculated using exact diagonalization (see $\S$\ref{sec:BYRO model} for details). The non-interacting limit is represented at $J_H/\zeta=0$. The highlighted region corresponds to  Yuan et al.'s chosen parameters for \BYRO{} \cite{BYRO_RIXS_Yuan2017}.\label{fig:fig1}}
\end{figure}

Fascination with unconventional materials that do not fit the conventional band theory of solids remains a core thread that connects condensed matter physics research past and present.
These so-called strongly correlated materials typically possess partially filled electronic bands and thus would conventionally be expected to be standard metals, however, they defy this fate and instead develop unconventional metallic or even insulating properties due to strong electronic correlations \cite{Correlation_effects1}.
While these concepts are no longer new, correlated metals and Mott insulators still garner significant attention due to the plethora of material archetypes with novel physical properties that they contain.
One example which has drawn extensive interest is the ordered double perovskites (DPs) of unit formula $\mathrm{A_2BB'O_6}$, which have gained tremendous popularity as a highly customizable playground to realize exotic and diverse physical properties \cite{DPs_review,5d1_DPs_Chen,5d2_DPs_Chen}.
Key to this customizability is their crystal structure, which takes the standard $\mathrm{ABO_3}$ perovskite structure and doubles its unit cell by replacing half of the B site transition metal (TM) ions with different $\mathrm{B'}$ TM ions.
The resulting crystal structure features a checkerboard pattern of corner-sharing $\mathrm{B(B')O_6}$ octahedra that surround alkali, alkaline-earth, or rare-earth $\mathrm{A}$ site ions, shown in Fig. \ref{fig:fig1a}.
The choice of B, $\mathrm{B'}$, and $\mathrm{A}$ site ions each has a profound impact on the resulting physical properties.
In fact, the diversity of DPs is such that widely varied states emerge even within the subset where the $\mathrm{A}$ site is restricted to be a non-magnetic alkaline-earth ion and the $\mathrm{B'}$ site to be a $5d^2$ ion.
Indeed, this subset includes phases as different as `hidden' multipolar-ordered Mott-insulating states, so dubbed because of their renowned difficulty to observe experimentally, and correlated high-$\mathrm{T_c}$ ferrimagnetic half-metals well suited for spintronic applications \cite{spintronics1,spintronics2,spintronics3,5d2_DPs_Chen,5d2_DPs_Paramekanti,5d2_DPs_voleti,Correlation_effects2}.

The first of these phases can be manifested when the B site possesses a closed electronic shell, leaving the $\mathrm{B'}$ site as the only electronically active ion.
This arrangement quashes nearest neighbour (NN) hopping between B and $\mathrm{B'}$ sites, leaving only next-neareast neighbour (NNN) hopping between $\mathrm{B'}$ sites.
The NNN hopping occurs across $\sqrt{2}$ times the NN distance and is consequently much weaker, promoting Mott-insulating phases \cite{5d2_DPs_Chen,Correlation_effects2}.
Within this Mott-insulating regime, then, we consider an atomic description of the electronic ground state and excited levels, starting from a $5d^1$ occupation following Refs. \cite{5d1_DPs_Chen,5d2_DPs_Chen,5d2_DPs_Paramekanti,5d2_DPs_voleti}.
The first and largest energy scale we consider is the crystalline electric field (CEF) exerted by the ligand oxygens, which in the ideal case of a cubic crystal structure splits the ten-fold degenerate $5d$ levels by an energy $\Delta$ into a lower energy \tg{} sextet and higher energy $e_g$ quartet.
Working in the \tg{} manifold, which possesses three-fold orbital degeneracy, it is convenient to describe its constituent microstates as possessing an effective orbital angular momentum \ltg{}.
This \leff{} description is subsequently converted to a \jeff{} description when spin-orbit coupling (SOC) is applied, splitting the \tg{} sextet into a ground state quartet carrying effective total angular momentum \jgs{} and an excited state doublet with \jes{}. 
In the large CEF limit, these levels are separated by $3/2$ times the SOC strength $\zeta$ and are inverse-ordered to maximize \jeff{} as a result of a sign change incurred within the \leff{} scheme.
More generally, $t_{2g} - e_g$ mixing weakly splits the \jgs{} levels and increases the separation between the ground state and the \jes{} doublet by $\frac{3}{2}\frac{\zeta^2}{\zeta/2+\Delta}$ to $\frac{3}{2}\zeta(1+\frac{\zeta}{\zeta/2+\Delta})=\frac{3}{2}\zeta_{eff}$, where $\Delta$ is the CEF strength \cite{soc_eff_t2g_eg_mixing}.
Previous studies have often used schemes disregarding the \eg{} orbitals ($\Delta=\infty$) which consequently enforces $\zeta=\zeta_{eff}$ \cite{HundsvsHubbard_metalvsins_Georges}.
In the case of two electrons occupying the spin-orbit coupled $5d$ levels, the introduction of electronic correlation, in the form of the Hund's coupling $J_H$, necessitates that we balance two limiting-case descriptions: i) a non-interacting system with $J_H/\zeta=0$ and ii) a strongly interacting one with $J_H/\zeta \gg 1$.

We begin with the large SOC limit in which two one-electron states are combined after SOC is considered, which thus connects directly to the $5d^1$ description above.
In the absence of Hund's coupling, these states can be combined in three ways: a ground state manifold with both electrons in a \jgs{} state, an excited manifold at $3\zeta_{eff}/2$ above the ground state with one electron in a \jgs{} state and the second in a \jes{} state, and an excited state at $3\zeta_{eff}$ above the ground state with both electrons  in a \jes{} state.
These combinations are illustrated graphically as transitions from the ground state configuration in Fig. \ref{fig:fig1b}.
The ground state manifold contains $\binom{4}{2}=6$ states, five with total effective angular momentum \Jtwo{} and one with $J_{eff}=0$, respectively described by the combined angular momentum terms symbols $^{2S+1}L_J=$$^{3}P_2$ and $^3P_0$.
The first excited manifold contains $4\times2=8$ states, five with \Jtwo{} ($^1D_2$) and three with $J_{eff}=1$ ($^3P_1$).
The second excited manifold contains $\binom{2}{2}=1$ state, which possesses $J_{eff}=0$ ($^1S_0$).

The second limit, with $J_H/\zeta \gg 1$, fittingly uses Hund's rules to determine the electronic ground state.
In order to impose dominance of Hund's coupling over SOC while simultaneously requiring $J_H<\Delta$, we pick up from the pre-SOC one-electron \ltg{} description of the \tg{} manifold.
Within the \tg{} manifold, Hund's rules dictate that the two-electron ground state possesses $S=1$ and \Ltg{}, and effective total angular momentum \Jtwo{} due to SOC, maximized as in the single-electron case. 
This ground state, the same $^{3}P_2$ manifold as in the non-interacting scheme, is stabilized compared to the lowest excited states (the $^{3}P_1$ triplet and $^{3}P_0$ singlet) by SOC, and from the subsequent $^1D_2$ quintet and $^1S_0$ singlet excited states by Hund's coupling in combination with SOC. 
The evolution of the $5d^2$ levels between these extrema is shown in Fig. \ref{fig:fig1c} as a function of $J_H/\zeta$, which reveals that the $^3P_2$ manifold is stabilized as the unique ground state for any $J_H>0$.
The resulting $^3P_2$ ground state phase space is very rich, supporting novel multipolar ordered states, including reduced moment dipolar, quadrupolar, and octupolar orders \cite{5d2_DPs_Chen,5d2_DPs_Paramekanti,5d2_DPs_voleti}. 
Indeed, recent measurements have revealed that \BYRO{} orders magnetically below $\mathrm{T_m\sim31}$ K with a heavily reduced ordered moment of $<0.5\ \mu_B$ and with strong indications of underlying quadrupolar order \cite{BYRO_order,BYRO_quadrupolar_order}.
Moreover, the two transitions observed in the specific heat at $\mathrm{T_q\sim37}$ K and $\mathrm{T_m\sim31}$ K match theoretical predictions of a two-step quadrupolar-to-dipolar ordering transition \cite{5d2_DPs_Chen,BYRO_quadrupolar_order,BYRO_synthesis_Aharen2010}.
Earlier work has sought to improve our understanding of these electronic ground state properties using resonant inelastic X-ray scattering (RIXS), establishing that \BYRO{} belongs in the intermediate Hund's coupling regime where $\order{J_H/\zeta}\sim 1$ \cite{BYRO_RIXS_Yuan2017,BYRO_RIXS_Paramekanti2018}.

The second aforementioned class of $5d^2$ DPs, the high-$\mathrm{T_c}$ ferrimagnetic half-metals, can be manifested in $3d-5d$ ordered DPs where an electronically active $3d$ ion resides on the B site.
Here, the $3d$ and $5d$ ions typically order magnetically well above room temperature, forming antiferromagnetically coupled ferromagnetic sublattices and leaving a remnant net moment (spin up by convention) due to a larger $3d$ ion magnetic moment \cite{spintronics3}. 
Unfortunately, $\mathrm{B/B'}$ anti-site disorder often significantly impacts their magnetic and electronic properties \cite{AS_disorder_effect,SCRO_metal_growth2002,SCRO_growth_Hauser2012,AS_disorder_effect_SCRO}, making it difficult to reach a clear understanding.
Therefore, the synthesis of fully ordered \SCRO{} thin films has rightly drawn attention and positioned \SCRO{} as a model $3d-5d$ DP system that can help us understand this class of materials more broadly \cite{SCRO_growth_Hauser2012,SCRO_APD2017,SCRO_moment_direction_Yuan2021}.

In \SCRO{}, the B sites are occupied by $3d^3$ Cr ions and the $\mathrm{B'}$ sites by $5d^2$ Re ions respectively possessing $S=3/2$ and $S=1$ due to Hund's coupling.
Given the antiferromagnetic coupling of the two moments, a net ferrimagnetic moment of 1 $\mu_B$ is anticipated within the ionic picture.
The half-metallicity of this archetype is also borne due to Hund's coupling, illustrated nicely in \SCRO{} as follows \cite{spintronics1,SCRO_2004_3d5dordering}.
Because the Cr site electron configuration is $3d^3$ and is also split into \tg{} and $e_g$ levels, the \tg{} spin-up levels are completely filled and the lowest energy empty state is a spin down \tg{} level.
On the other hand, because the Re site configuration is $5d^2$, the spin-down levels are only partially filled and the lowest energy empty state is the remaining spin down \tg{} level due to Hund's coupling.
This means that the minority down spins may hop between Re and Cr sites and become itinerant while the majority up spins become localized on the Cr sites.
Correlation effects thus induce near complete spin-polarization at the fermi level, such that this archetype of materials is highly promising for spintronic applications \cite{spintronics1,spintronics2,spintronics3}.
Experimentally, ferrimagnetic ordering in \SCRO{} occurs below $\mathrm{T_c=}$ 508 K in fully ordered thin films with a saturated moment of $\sim 1.29\ \mu_B$ \cite{SCRO_growth_Hauser2012}.
This enhancement from the ionic picture prediction of 1 $\mu_B$ has been attributed to the presence of strong SOC, which theoretically should lead to a saturated moment of 1.28 $\mu_B$ \cite{SCRO_pseudo_half_metal_2005}.
Supporting the notion of strong SOC in \SCRO{}, which further promotes Mott-insulating phases, updated measurements of fully ordered \SCRO{} reveal that it is in fact a small band gap semiconductor rather than a true half-metal \cite{SCRO_pseudo_half_metal_2005,SCRO_growth_Hauser2012,HundsvsHubbard_3d35d3_Meetei,HundsvsSOC_metalvsins_gangchen}.
Moreover, recent RIXS measurements of \SCRO{} reveal features believed to represent Hund's coupling spin-orbit excitations, pointing to both strong SOC and Hund's coupling \cite{SCRO_RIXS}.

In this paper, we present Re $L_2$ and $L_3$ edge RIXS measurements of \SCRO{} and \BYRO{} that reveal intermediate Hund's coupling and strong SOC in both materials.
Moreover, we use the contrast between $L_2$ and $L_3$ edge RIXS spectra to directly identify the SOC features in these materials by way of single-particle RIXS selection rules which forbid SOC excitations at the $L_2$ edge.
Remarkably, the $L_2-L_3$ edge contrast also reveals the crossover from a single-particle description to a many-body description as correlation effects partially relax this strict selection rule and SOC excitations appear, though significantly suppressed, at the $L_2$ edge.
In \BYRO{}, this $L_2$ edge suppression of SOC features furthermore reveals that the order of the $^3P_1$ and $^3P_0$ levels is inverted relative to earlier parametrizations, necessitating a description with decreased SOC and increased Hund's coupling \cite{BYRO_RIXS_Yuan2017,BYRO_RIXS_Paramekanti2018}.
In order to refine the specific energy scales of the SOC and Hund's coupling we employed exact diagonalization methods to calculate the $L_2$ and $L_3$ edge RIXS spectra of \BYRO{} using the EDRIXS program.
We report that values $\zeta=0.290(5)$ eV and $J_H/\zeta=1.30(5)\ (J_H=0.38(2)\ \mathrm{eV})$ successfully reproduce the energy scales and spectral weight characteristics of the $L_2$ and $L_3$ edge RIXS features in \BYRO{}.
This newly refined value of $\zeta$ is more in line with that found in other Re-based Mott insulators \cite{Frontini_AMRO_2024,Ba2CaReO6_5d1_vibronic,5d3_K2ReCl6_SOC}.
Our measurements of \BYRO{} additionally reveal low energy features plausibly explained by the presence of a dynamic Jahn-Teller effect either in isolation or in conjunction with an $e-h$ continuum.
The presence of a dynamic Jahn-Teller effect could moreover help to explain the asymmetric, edge dependent lineshape of the atomic multiplet excitations \cite{5d2_vibronic_iwahara}.
In \SCRO{}, our measurements reveal the presence of previously unresolved excitations characteristic of intermediate Hund's coupling \cite{SCRO_RIXS}.
Moreover, we observe a strong $e-h$ continuum at energy transfer under $\sim 2.5$ eV which was also not discerned in previous studies and which may explain the low energy feature previously attributed to single-magnon excitations \cite{SCRO_RIXS}.
The $e-h$ continuum also presents with no discernible gap, reflecting \SCRO{}'s small band gap nature and furthermore pointing to an indirect gap in line with resistivity estimates and much smaller than optical estimates of the direct gap \cite{SCRO_growth_Hauser2012}.
Finally, a comparison of the RIXS spectra of \BYRO{} and \SCRO{} reveals similar energy scales of Hund's coupling and SOC despite their respective Mott-insulating and semiconducting natures, reflecting the crucial role of B site substitution in $5d^2$ DPs.
\section{\label{sec:experiment}Experimental Details}
Our experiments were performed with a high quality powder \BYRO{} sample and an unstrained 319 nm thick \SCRO{} thin film sample grown on a $\mathrm{SrTiO_3}$ substrate.
The growth and characterization of \BYRO{} and \SCRO{} have been previously reported \cite{BYRO_synthesis_Aharen2010,SCRO_growth_Hauser2012}.

RIXS measurements were conducted at the Advanced Photon Source at the 27-ID-B beamline.
The experiments were performed at the Re $L_2$ and $L_3$ X-ray absorption edges ($2p_{1/2}\to 5d$, $E_i=$ 11.960 keV and $2p_{3/2}\to 5d$, $E_i=$ 10.531 keV, respectively). 
The incident optics consist of a high heat load monochromator and a high resolution monochromator used to select the incident photon energy as well as a set of focusing optics to focus the beam at the sample surface.
The high heat load monochromator is a two-bounce diamond (111) crystal monochromator, while the high resolution monochromator is a four-bounce $\mathrm{Si}$ crystal monochromator with variable reflection condition. 
The four-bounce arangement is chosen so that the beam position does not change with a change in incident energy.
The experiment utilized the Si(440) reflection condition of the high resolution monochromator.
The incident energy was calibrated by measuring the total fluorescence yield X-ray absorption (TFY-XAS) of the powder \BYRO{} sample using a pin diode and setting the inflection point of the absorption to a reference value.

The receiving optics are arranged in the typical Rowland circle geometry to discriminate between emitted photon energies, with a diced, spherically-bent crystal analyzer of bending radius 2 m  and a Spectrum Lambda60k 2-dimensional pixelated strip detector.
The crystal analyzer used for the Re $L_3$ edge measurements had a Si(119) reflection condition, while the one used at the Re $L_2$ edge had a Si(773) condition.

In order to minimize the elastic background due to Thomson scattering, the experiments were conducted with horizontally polarized light in a horizontal four-circle geometry (ensuring polarization $\pi\to\pi',\sigma'$) with $2\theta$ close to $90^\circ$. 
The total energy resolution was $\sim$ 95 meV (FWHM) at the Re $L_2$ edge and $\sim$ 65 meV at the Re $L_3$ edge.
A closed cycle cryostat was used for temperature control.
The \SCRO{} sample was aligned with the (HHL) crystallographic plane in the scattering plane.
Because the thickness of the thin-film \SCRO{} sample is significantly less than the X-ray penetration depth, measurements were conducted at grazing incidence ($\theta = 1^\circ$) to increase the sample volume probed.
This helps maximize the RIXS signal to noise ratio.
\section{\label{sec:BYRO results}\secBYRO{} Results}
\subsection{\label{sec:BYRO common}General features }
\begin{figure}[h]
\includegraphics[width=\columnwidth]{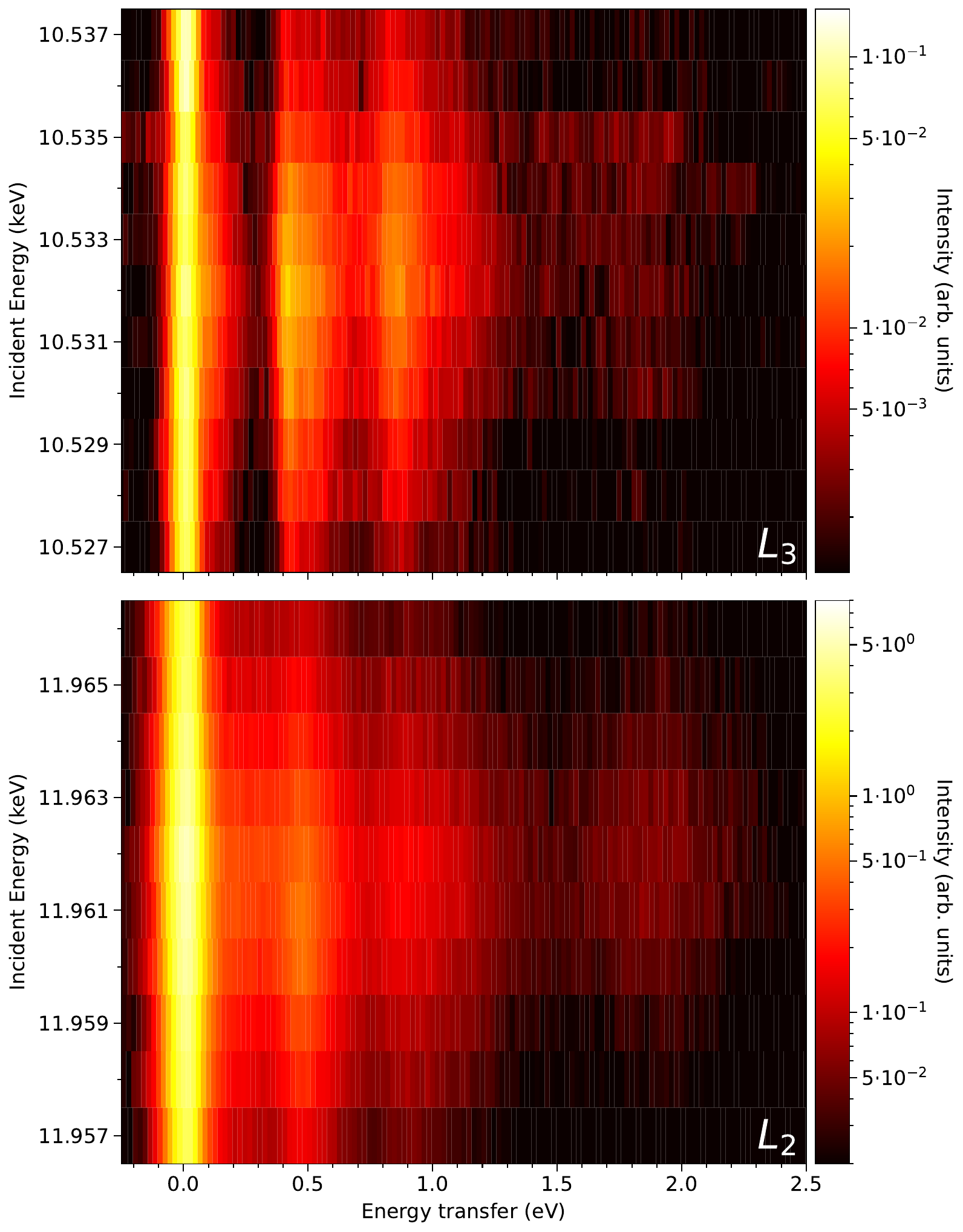}
\caption{Top (bottom): \BYRO{} Re $L_3$ ($L_2$) edge RIXS map with $\mathrm{E_t}$ up to 2.5 eV collected at $\mathrm{T=18}$ K.\label{fig:fig2}}
\end{figure}

RIXS maps of excitations in \BYRO{} up to 2.5 eV in energy transfer ($\mathrm{E_t}$) are shown in Fig. \ref{fig:fig2}.
The Re $L_2$ and $L_3$ edge RIXS maps were measured at a temperature of $\mathrm{T=18}$ K.
Limited measurements as a function of temperature at the Re $L_2$ edge do not show significant temperature dependence, see the Supplemental Material \cite{supp} (see also Refs. \cite{Cowan_Theory_1981,konig_nephelauxetic_1971} therein).
The two RIXS maps display the same primary features in addition to the elastic line: two strong features at $\sim$ 0.5 eV and $\sim$ 0.9 eV and a third, weaker feature at $\sim$ 2 eV. 
The Re $L_2$ data have higher statistics but the features are also visually broader than their Re $L_3$ counterparts due to the coarser resolution.
The sub 1 eV features resonate at an incident photon energy of 10.532-10.533 keV at the Re $L_3$ edge and at 11.961 keV at the Re $L_2$ edge. 
In comparison, the 2 eV feature resonates at a slightly higher incident photon energy of 11.962 keV at the Re $L_2$ edge, however this trend is unfortunately not distinguishable at the Re $L_3$ edge due to the poorer statistics.
The higher incident energy resonance of the excitation indicates that the feature represents a $dd$ excitation, i.e. an excitation in which the incident photon promotes a core electron to an excited $5d$ state and a ground state electron de-excites to annihilate the induced core-hole.
The end result is the transition of an electron from the ground state to an excited $5d$ state, and this process resonates at an incident photon energy that increases linearly with the energy of the excited state.
Thus, a 2 eV $dd$ excitation will resonate at an $\mathrm{E_i}$ 1 eV above the resonance of a 1 eV $dd$ excitation, precisely as we have observed. 
The energy scale of the features and identification of these features as $dd$ excitations generally agrees with prior Re $L_2$ and $L_3$ edge RIXS measurements and theory of \BYRO{} shown in Fig. \ref{fig:fig1c} \cite{BYRO_RIXS_Yuan2017,BYRO_RIXS_Paramekanti2018}.
These theoretical calculations identified the 0.5 eV feature as a doublet excitation composed of one sub-level ($^3P_0$) primarily reflective of the Hund's coupling $J_H$ and a second sub-level ($^3P_1$) primarily reflective of the SOC $\zeta$, shown schematically in Fig. \ref{fig:fig1b}.
Meanwhile, the  0.9 eV feature was theorized to reflect $J_H$ and $\zeta$ combined.
Finally, the 2 eV feature was identified as a `two-particle' SOC excitation to the $^1S_0$ level of nominal energy $3\zeta_{eff}$ which increases with $J_H$.
This excitation to the $^1S_0$ level is dubbed a `two-particle' excitation because it corresponds to an excitation of two electrons from the \jgs{} manifold to the \jes{} doublet from a single-particle perspective, though it is more accurately described as occurring due to correlation-driven mixing of the \jgs{} and \jes{} levels in the multi-particle regime with $J_H>0$ \cite{BYRO_RIXS_Yuan2017,BYRO_RIXS_Paramekanti2018}.
The low spectral weight of the 2 eV feature is also in agreement with prior results and theory, which predicts the intensity of the `two-particle' excitations to be suppressed compared to the single-particle excitations \cite{BYRO_RIXS_Yuan2017,BYRO_RIXS_Paramekanti2018}.

\begin{figure}[h]
\includegraphics[width=\columnwidth]{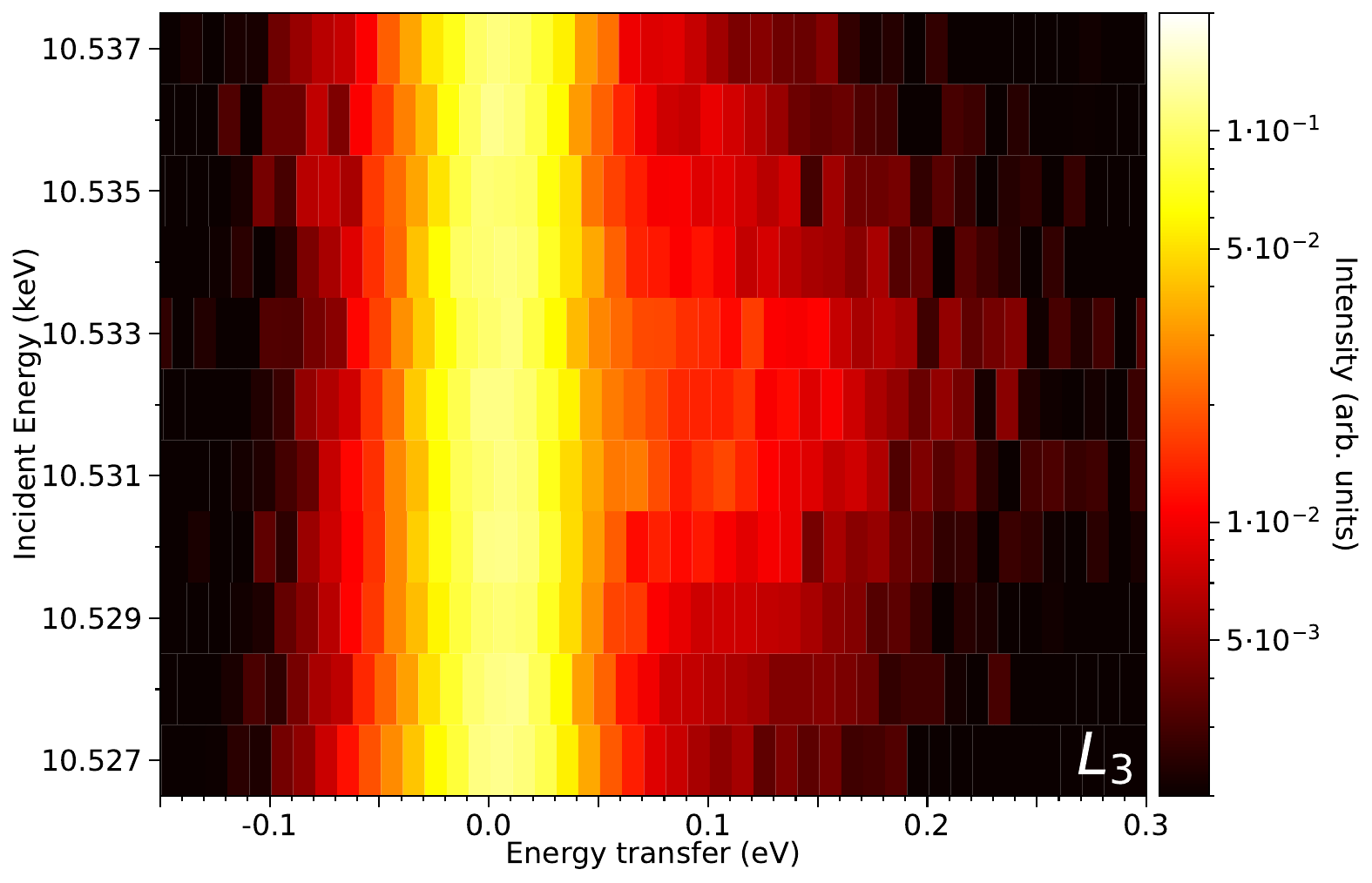}
\caption{Re $L_3$ edge RIXS map of \BYRO{} with $\mathrm{E_t}$ up to 0.3 eV obtained at $\mathrm{T=18}$ K.\label{fig:fig3}}
\end{figure}

Besides these three primary features, and not as obvious in Fig. \ref{fig:fig2}, is a resonant low energy excitation that presents as a shoulder in the quasi-elastic line at both edges, which was not resolved in the aforementioned prior measurements.
In order to better resolve this low energy feature we collected a RIXS map of the quasi-elastic region at the Re $L_3$ edge at $\mathrm{T=18}$ K, shown in Fig. \ref{fig:fig3}.
The increased detail of this RIXS map clearly shows the presence of a resonant low energy excitation that extends up to $\mathrm{E_t}$ of $\sim$ 0.25 eV.
The maximum of resonance occurs at an incident photon energy of $\mathrm{E_i}\sim$ 10.532-10.533 keV, matching the excitations at 0.5 eV and 0.9 eV, indicating that this feature either represents a low energy $dd$ excitation, implying a splitting of the \Jtwo{} ground state, or represents some low energy non-$dd$ excitation of the ground state.
The possible origins of this feature are discussed at greater length in $\S$\ref{sec:discussion}.

\subsection{\label{sec:BYRO edge dep}\secLII{} and \secLIII{} edge comparison}
\begin{figure}[h]
\includegraphics[width=\columnwidth]{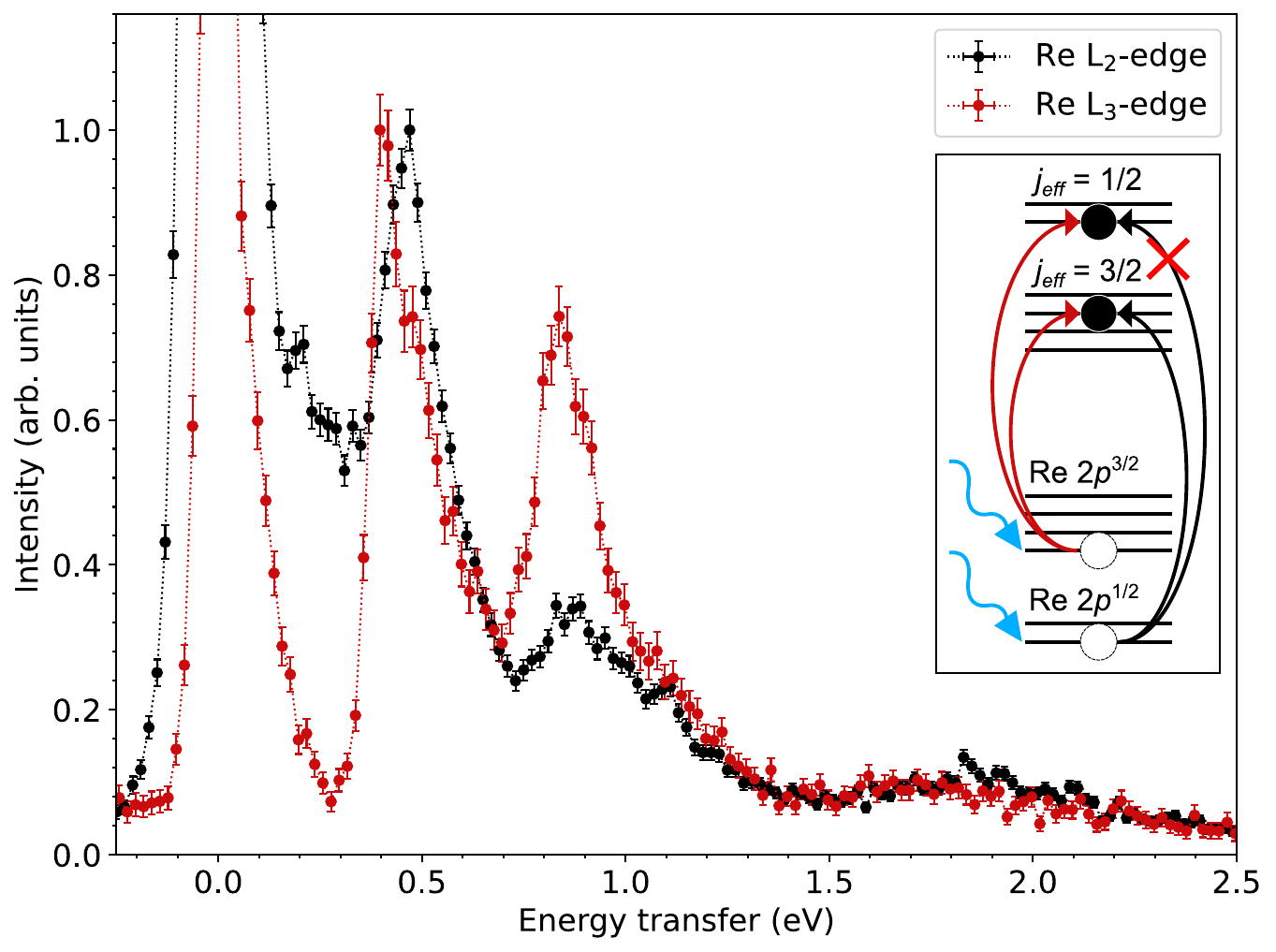}
\caption{\BYRO{} RIXS spectra at the Re $L_2$($L_3$) edge resonances up to 2.5 eV in $\mathrm{E_t}$ collected at $\mathrm{T=18}$ K and $\mathrm{E_i}=11.961$(10.533) keV. The RIXS spectra are normalized by the maximum intensity of the 0.5 eV feature. The right inset schematically shows the allowed and forbidden single-particle RIXS transitions at the $L_2$ and $L_3$ edges. \label{fig:fig4}}
\end{figure}

In order to gain a better understanding of the excitations in \BYRO{} we compare the Re $L_2$ and $L_3$ edge spectra obtained at resonance ($\mathrm{E_i}=11.961$ keV and 10.533 keV), shown in Fig. \ref{fig:fig4} normalized by the 0.5 eV feature maximum.
Normalization by the 0.5 eV feature is done for easier comparison as the 0.5 eV and 0.9 eV features are much weaker relative to the elastic line at the $L_2$ edge, a phenomenon which relates to the overall differences between $L_2$ and $L_3$ edges which will be discussed shortly.
This closer comparison of the Re $L_2$ and $L_3$ edge spectra reveals notable differences in the spectra even after the overall scale difference is accounted for.

The first and most stark difference is a shift in the maximum of the first feature, with the Re $L_3$ data peaked around 0.41 eV and the Re $L_2$ data peaked around 0.47 eV.
This difference becomes more remarkable when we analyze the physical meaning of the excitations in the spectra.
As discussed in $\S$\ref{sec:intro}, the $5d^2$ ground state for intermediate Hund's coupling is the five-fold degenerate $^3P_2$ manifold, which is stabilized compared to the $^3P_1$, $^3P_0$, $^1D_2$, and $^1S_0$ excited levels by a combination of Hund's coupling and SOC. 
The energy levels of these excited states relative to the ground state depends on both $\zeta$ and $J_H/\zeta$ as shown in Fig. \ref{fig:fig1c}.
The levels evolve fairly smoothly and with preserved hierarchy as a function of $J_H/\zeta$, with the notable exception being the $^3P_1$ and $^3P_0$ levels.
These states approach each other, forming the doublet peak observed by Yuan et al., and ultimately cross around $J_H/\zeta =1.1$ before the $^3P_0$ state further elevates in the large $J_H$ limit.
Yuan et al. modelled the spectra using a Kanamori parametrization restricted to the \tg{} levels, such that $\Delta=\infty$ and $\zeta=\zeta_{eff}$, and conclude the Kanamori $J_H^{eff}\ (\sim 5/7 J_H)$ to be 0.26 eV and the \tg{} $\zeta_{eff}$ to be 0.38 eV \cite{HundsvsHubbard_metalvsins_Georges}.
Within a full $d$ level parametrization with a physical CEF strength of $\Delta\sim4.5$ eV these values become $J_H\sim0.34$ eV and $\zeta\sim0.35$ eV, giving a ratio of $J_H/\zeta\sim0.95$, that places the $^3P_0$ level below the $^3P_1$ level as highlighted in Fig. \ref{fig:fig1c}.
The differences in the $L_2$ and $L_3$ edge spectra, however, tell a different story when $L_2$ and $L_3$ edge selection rules are considered.
As discussed in $\S$\ref{sec:intro} and as illustrated in Fig. \ref{fig:fig1b}, from a single-particle perspective a transition to the $^3P_1$ level corresponds to promoting one electron from the \jgs{} level to the \jes{} level, which is derived from $J=5/2$ as illustrated in the study of iridates \cite{L2_selection_rules_iridium_Ament,L2_selection_rules_iridium_Clancy}.
Thus, the atomic \jes{} level cannot be probed at the $L_2$ edge since this would require a dipole forbidden transition with $\Delta J=2\ (2p_{1/2}\to5d_{5/2})$, shown schematically in the inset of Fig. \ref{fig:fig4}.
While the strict selection rule is relaxed by the introduction of Hund's coupling, we nevertheless anticipate this transition to be suppressed at the $L_2$ edge.
Thus, if the $^3P_0$ state were to lie lower in energy than the $^3P_1$ manifold we would expect the $L_2$ RIXS spectrum to have a lower spectral weight in the high energy region of the doublet feature, whereas the opposite trend is clearly observed.
It stands to reason then that the spectral differences between the $L_2$ and $L_3$ edges inform us that the $^3P_1$ manifold lies below the $^3P_0$ state, i.e. $J_H/\zeta \gtrsim 1.1$ based on Fig. \ref{fig:fig1c}.
Moreover, this increase in $J_H$ necessitates a decrease in $\zeta$ from 0.35 eV to maintain the overall excited state energy levels, which would in fact put it more in line with other Re-based Mott insulators that display $\zeta \sim0.3$ eV \cite{Frontini_AMRO_2024,Ba2CaReO6_5d1_vibronic,5d3_K2ReCl6_SOC}.

With the edge dependence trend of the doublet feature explained, we now apply this single-particle selection rule lens to other differences between the $L_2$ and $L_3$ edge spectra.
To this end we start with the aforementioned question of overall scale, i.e. why the $L_2$ edge features are so much weaker relative to the elastic.
This follows easily using the same argument as for the $^3P_1$ manifold: both the 0.9 eV transition to the $^1D_2$ level and 1.8 eV (refined from the rough estimate of 2 eV from the RIXS map) transition to the $^1S_0$ level involve the $L_2$-forbidden single-particle \jes{} state, and will thus also be suppressed compared to the $L_3$ edge.
Next, we turn to the suppression of the 0.9 eV feature at the $L_2$ edge beyond the overall scale factor.
Its suppression compared to the 0.5 eV feature can be naturally explained: whereas the 0.5 eV feature contains the $L_2$-allowed $^3P_0$ sub-level, the 0.9 eV feature is only composed of the $L_2$-forbidden $^1D_2$ level and should therefore be more suppressed compared to the doublet 0.5 eV feature.
On the other hand, its suppression compared to the 1.8 eV $^1S_0$ feature is less easy to reconcile, given that it is also solely composed of $L_2$ suppressed transitions.
Speculatively, the answer may lie in the fact that the $^1S_0$ transition is, as aforementioned, a `two-particle' excitation mediated by correlation effects, weakening the validity of edge-based selection rules taken from the single-particle picture.
Indeed, the `two-particle' excitation nature of the $^1S_0$ feature appears to be manifest in the fact that it is much broader than the single-particle excitation features, which we interpret as lifetime broadening that indicates the unstable nature of the `two-particle' excitation in the final state.
Overall, then, the application of single-particle selection rules does an impressive job in explaining the $L_2-L_3$ edge discrepancies of the atomic $dd$ excitations within the standard \Jtwo{} ground state theory.
While this general description indeed fits our results, we are interested to provide a more quantitative one below.

\subsection{\label{sec:BYRO model}RIXS calculations}
\begin{figure}[b]
\includegraphics[width=\columnwidth]{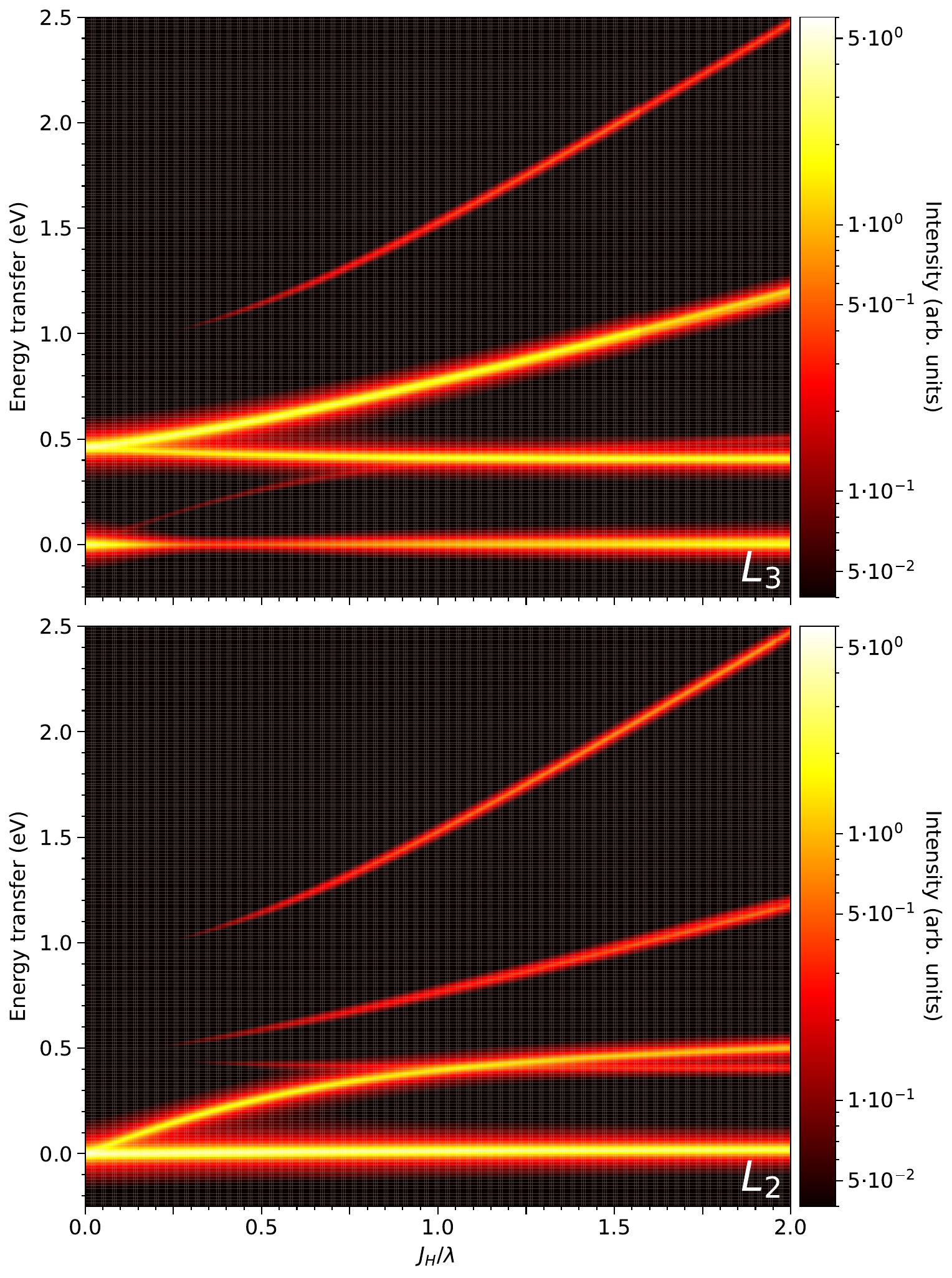}
\caption{Top (bottom): EDRIXS calculated Re $L_3(L_2)$ edge RIXS heatmap of a $5d^2$ system as a function of $J_H/\zeta$ and $\mathrm{E_t}$. The heatmaps are calculated for fixed $\zeta=0.29$ eV at a temperature T = 18 K.\label{fig:fig5}}
\end{figure}
In order to quantitatively estimate the energy scales of $J_H$ and $\zeta$, we calculate the Re $L_2$ and $L_3$ edge RIXS spectra of \BYRO{} with exact diagonalization computational methods using the EDRIXS program \cite{EDRIXS_WANG2019}.
In order to refine appropriate values of $J_H$ and $\zeta$, we calculated the RIXS spectra of a $5d^2$ ion at both Re $L_3$ and $L_2$ edges for variable $\zeta\in[0.25,0.33]$ eV and $J_H/\zeta\in[0,2]$, and fixed $\Delta=4.5$ eV (see $\S$\ref{sec:SCRO high E}, Ref. \cite{BYRO_RIXS_Yuan2017}).
The calculated spectra were measured against the experimental spectra, resulting in optimized values of $\zeta=0.290(5)$ eV and $J_H/\zeta=1.30(5)$.
The optimization process and other details of the EDRIXS calculations are discussed in the Supplemental Material \cite{supp}.

\begin{figure}[h]
\includegraphics[width=\columnwidth]{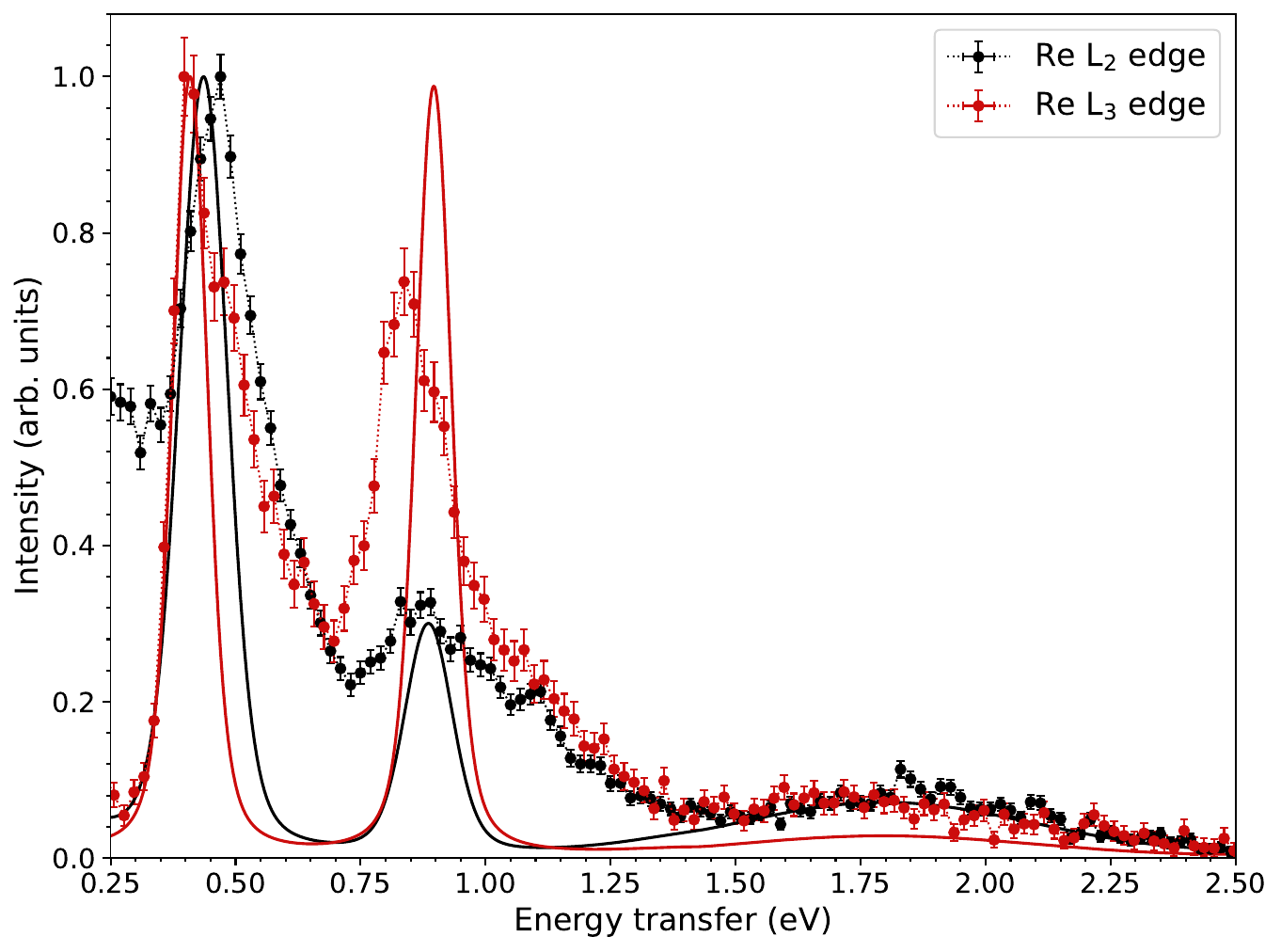}
\caption{Experimental and EDRIXS calculated Re $L_3$ (red) and $L_2$ (black) edge RIXS spectra of \BYRO{} at T = 18 K. The spectra are calculated with the optimized values of  $J_H/\zeta =1.30$ and $\zeta=0.290$ eV.\label{fig:fig6}}
\end{figure}

The calculated RIXS spectra with $\zeta=0.29$ eV are shown in heatmap form as a function of $J_H/\zeta$ and $\mathrm{E_t}$ in Fig. \ref{fig:fig5} with intensity corresponding to the RIXS signal.
The spectra are only moderately broadened to 15 meV FWHM to more clearly show the sub-levels of the $^3P_0+^3P_1$ doublet feature.
Fig. \ref{fig:fig5} clearly replicates the Hund's coupling dependent trends of the atomic energy levels shown in Fig. \ref{fig:fig1c} with added information of the edge-dependent RIXS signal intensity.
The intensity profile of the spectra reflect and substantiate several key points discussed in $\S$\ref{sec:BYRO edge dep}.
First, we observe that at both edges the intensity of the `two-particle' $^1S_0$ excitation disappears in the non-interacting limit of $J_H/\zeta=0$ as anticipated.
Next, we see that the $^3P_1$ and $^1D_2$ SOC excitations also vanish in this limit at the $L_2$ edge in correspondence with the aforementioned selection rule in the non-interacting case.
Finally, we observe that this selection rule is gradually relaxed as a function of $J_H/\zeta$ but that these excitations remain suppressed at the $L_2$ edge compared to the $L_3$ edge, reflecting a smooth crossover from single-particle to many-body physics.
Correspondingly, the calculation also corroborates our earlier speculation that the $\sim0.5$ eV doublet would be spectrally biased to higher energy at the Re $L_3$ edge for $J_H/\zeta\lesssim1.1$ and at the Re $L_2$ edge for $J_H/\zeta\gtrsim1.1$.
This is reflected in the optimized value of $J_H/\zeta=1.30(5)$, for which we show the RIXS spectra in Fig. \ref{fig:fig6}.

The optimized spectra shown in Fig. \ref{fig:fig6} are normalized by the $^3P_0+^3P_1$ doublet intensity and are broadened to the instrumental resolutions, with the exception of the $^1S_0$ excitation.
Since this excitation has a ‘two-particle’ origin in the single particle picture, we expect that it may be unstable (more so than other $dd$ excitations), showing  the lifetime broadening effect.
Fig. \ref{fig:fig6} reveals good overall agreement between the EDRIXS calculations and the experimental spectra both in terms of the energy levels of the excitations and their relative intensities at each edge.
In particular, the calculated spectra reproduce the shifting of the $^3P_0+^3P_1$ doublet to higher energy from the $L_3$ to $L_2$ edge, the greater suppression of the $^1D_2$ excitation at the $L_2$ edge compared to the other two features, and the low intensity of the $^1S_0$ excitation at both edges.
Despite this broad agreement, we take note of the fact that the shifting of the $^3P_0 + ^3P_1$ doublet is underestimated for this value of $J_H/\zeta$ and that the $^1D_2$ excitation energy is slightly overestimated.
These characteristics are not independent and reflect a balance that must be struck to achieve the greatest overall agreement, where the shifting of the doublet can be matched by increasing $J_H/\zeta$ but at the direct cost of further overestimating the energy of the $^1D_2$ transition and additionally overestimating the energy of the $^1S_0$ transition.
The reason for this slight quantitative deviation from the experimental results is not well understood, but could for instance be caused by the energy-level-shifting vibronic coupling effects seen in several related Re-based DPs and whose suspected presence in \BYRO{} is further discussed in $\S$\ref{sec:discussion} \cite{Frontini_AMRO_2024,Ba2CaReO6_5d1_vibronic,5d2_vibronic_iwahara}.
Overall, the atomic EDRIXS calculations do a good job in qualitatively and quantitatively replicating the RIXS spectra of \BYRO{} and, moreover, demonstrate the applicability of our preceding single-particle selection-rule based analysis.

\section{\label{sec:SCRO results}\secSCRO{} results}

\subsection{\label{sec:SCRO high E}High energy features}
\begin{figure}[h]
\includegraphics[width=\columnwidth]{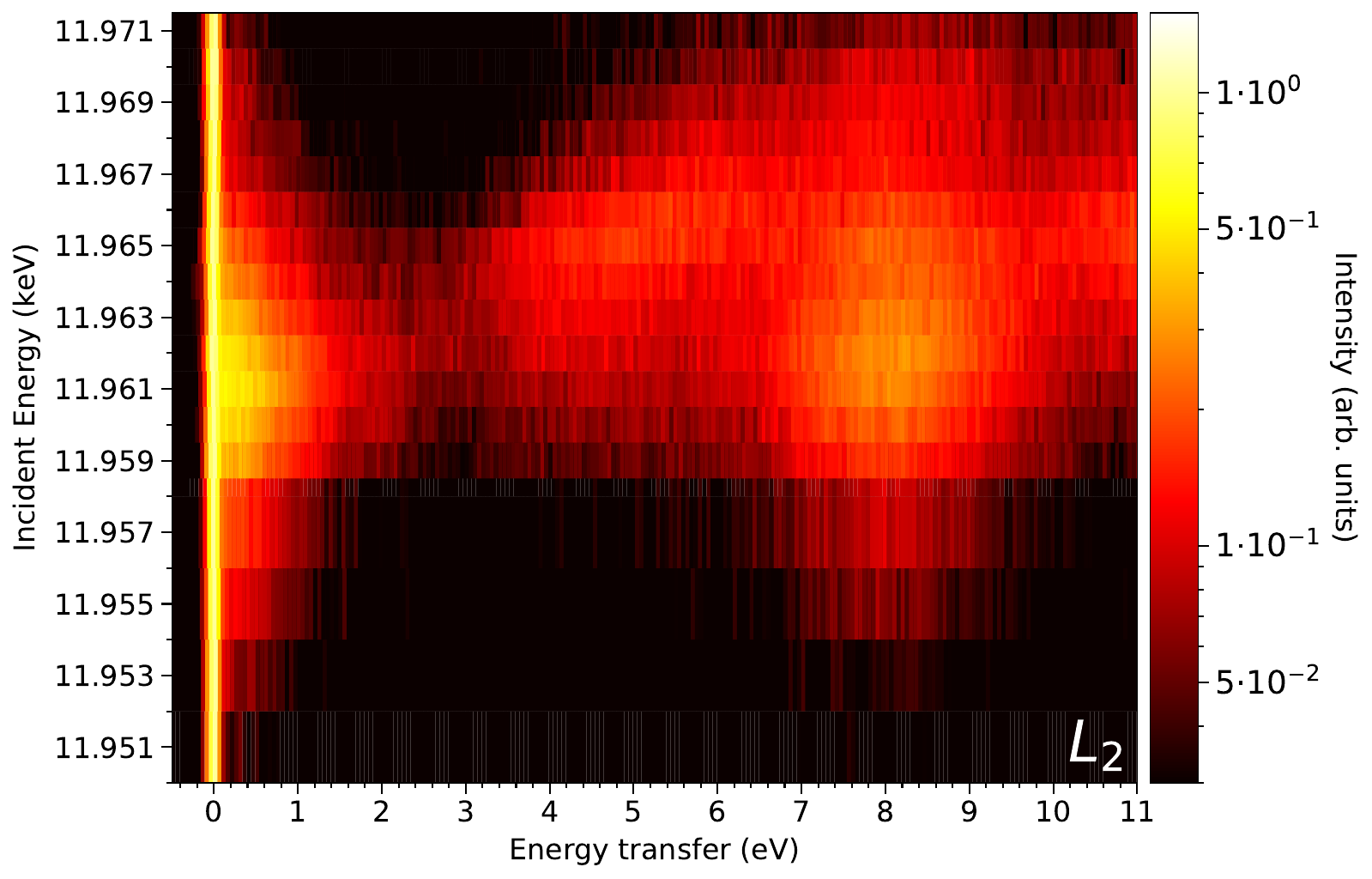}
\caption{\SCRO{} Re $L_2$ edge RIXS map up to 11 eV in $\mathrm{E_t}$. The spectra were collected at $\mathrm{T=300}$ K and are normalized by the elastic line.\label{fig:fig7}}
\end{figure}

A RIXS map of the \SCRO{} excitations up to 11 eV was measured at the Re $L_2$ edge at $\mathrm{T=300}$ K, shown in Fig. \ref{fig:fig7}. 
Three broad features are observed in Fig. \ref{fig:fig7}: a continuum extending from the elastic line up to $\sim$ 2.5 eV in $\mathrm{E_t}$, a feature centered at $\sim$ 4.5 eV in $\mathrm{E_t}$, and a feature centered at $\sim$ 8 eV in $\mathrm{E_t}$.
The first of the features resonates at $\mathrm{E_i}$=11.961 keV and appears to decrease monotonically in intensity with $\mathrm{E_t}$.
The resonance energy of this feature and its extended nature likely indicate that it corresponds to excitations to the conduction levels directly above the ground state, i.e. into the electron-hole ($e-h$) continuum.
Indeed, similar excitations were previously observed and so-diagnosed in the related $3d-5d$ DPs \BFRO{} and \CFRO{} \cite{BYRO_RIXS_Yuan2017}.
The breadth of the feature may therefore give an rough measure of the conduction band width and points to a small or non-existent band gap, both in line with the small band gap semiconducting nature of \SCRO{}.
The second of the aforementioned features resonates at a higher incident photon energy of $\mathrm{E_i}$=11.965-11.966 keV, marking it as $dd$ excitation as described in $\S$\ref{sec:BYRO common}.
The energy of the excitation further identifies it as the transition between the \tg{} ground state and $e_g$ excited states, whose nominal energy $\Delta\sim4.5$ eV defines the ligand CEF strength.
Contrastingly, the third feature resonates at the same incident photon energy as the low energy $e-h$ continuum, marking it instead as charge-transfer excitations from the ligand oxygens rather than a $dd$ excitation.
Charge-transfer excitations refer to the process in which hybridized Re$-$O states (O $2p$ states in the atomic picture) de-excite to annihilate the Re site core-hole during the RIXS process.
The lowest energy charge-transfer excitations share the same $t_{2g}^3$ intermediate state as the lowest energy $dd$ excitations and $e-h$ continuum, causing them to resonate at the same incident photon energy; however, they occur at a larger energy transfer because the $t_{2g}^3 \bar{L}$ final state featuring a ligand hole is more energetic than the corresponding $t_{2g}^2\ dd$ excitation/ $e-h$ continuum final state.
A hint of a fourth feature is also visible above 10 eV in $\mathrm{E_t}$ and extending beyond the limit of 11 eV.
The feature appears to resonate at the same energy as the $e_g$ feature, marking it as a charge-transfer + crystal field excitation in which an electron is excited to an empty $e_g$ state and a ligand oxygen electron de-excites to annihilate the core hole.

\subsection{\label{sec:SCRO low E}Low energy features}
\begin{figure}[h]
\includegraphics[width=\columnwidth]{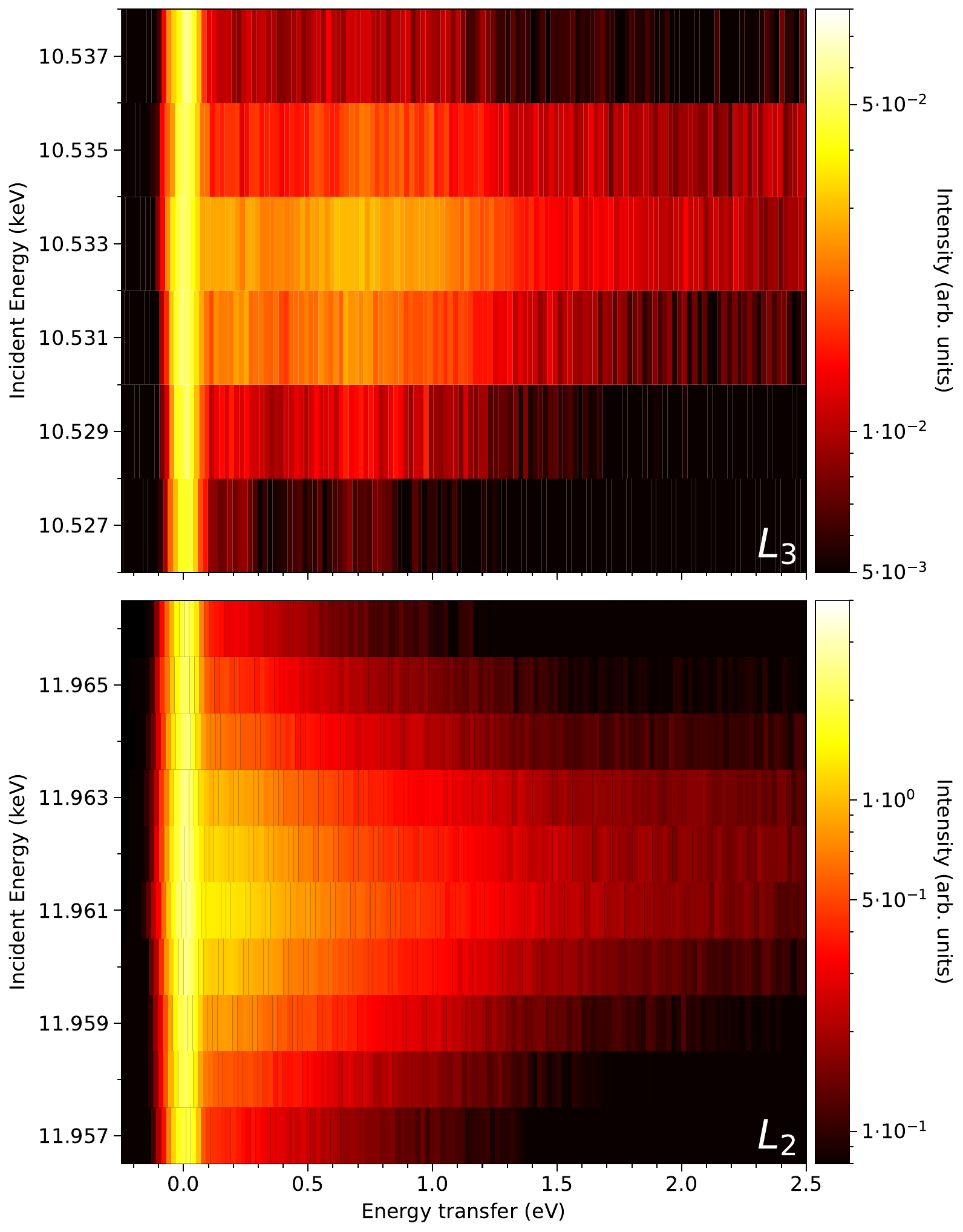}
\caption{Top (bottom): \SCRO{} Re $L_3$($L_2$) edge RIXS map up to 2.5 eV in $\mathrm{E_t}$. The data were collected at $\mathrm{T=300}$ K.\label{fig:fig8}}
\end{figure}

\begin{figure}[h]
\includegraphics[width=\columnwidth]{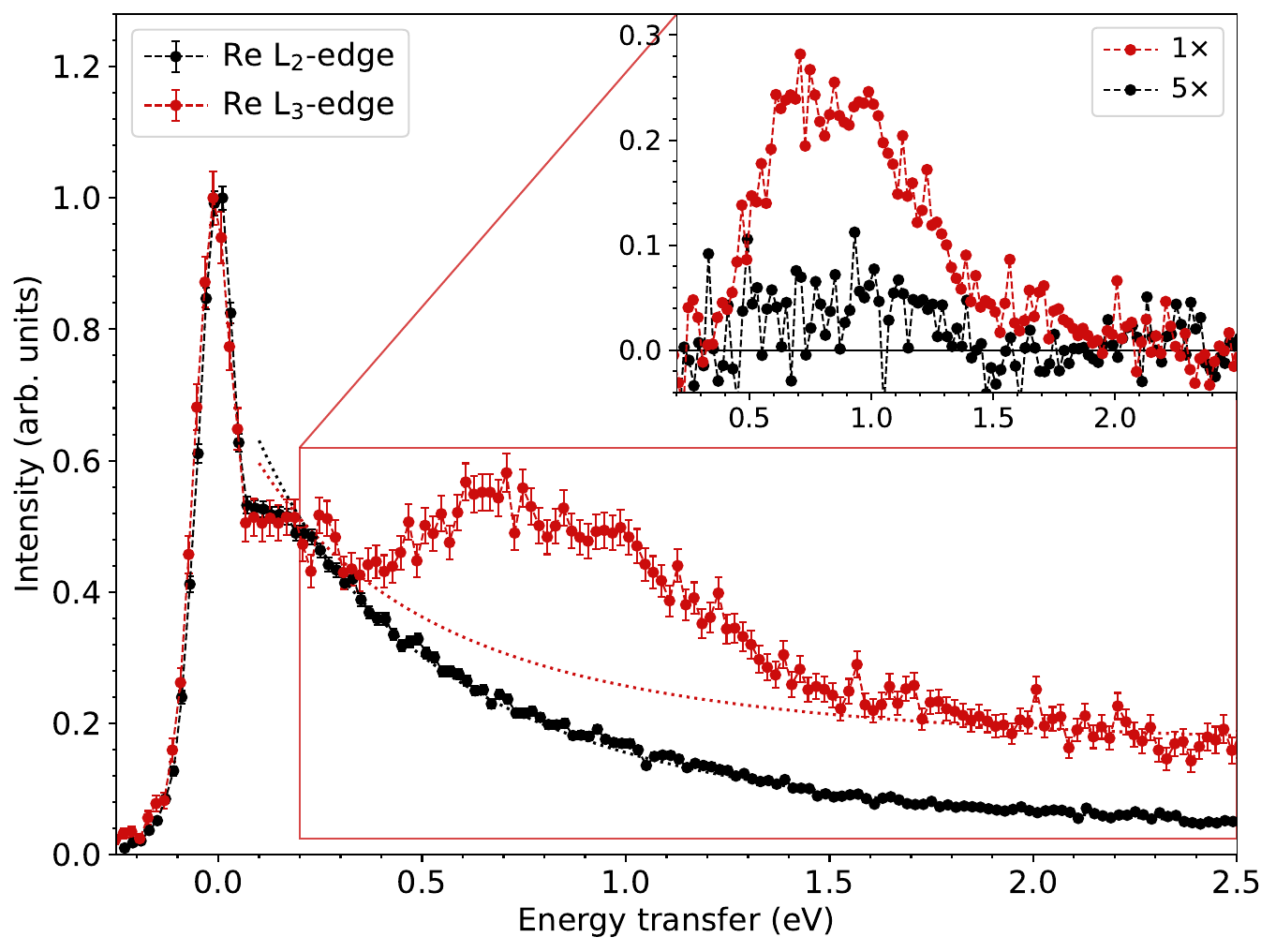}
\caption{\SCRO{} RIXS spectra at the Re $L_2$($L_3$) edge resonance of $\mathrm{E_i}=11.961$(10.533) keV up to 2.5 eV $\mathrm{E_t}$. The data are overlaid with the $L_2$ and $L_3$ edge decaying power law fits of the $e-h$ continuum, respectively indicated by the black and red dashed lines. The zoomed inset shows the spectra with the $e-h$ continuum background subtracted. The data were collected at $\mathrm{T=300}$ K and are normalized by the elastic line intensity for greater comparability. \label{fig:fig9}}
\end{figure}

RIXS maps of the low energy continuum in \SCRO{} were measured with greater point density at the Re $L_2$ and $L_3$ edges at $\mathrm{T=300}$ K, shown in Fig. \ref{fig:fig8}.
The RIXS maps in Fig. \ref{fig:fig8} both show the aforementioned continuum, however, there is a clear discrepancy between $L_2$ and $L_3$ edges.
The $L_3$ RIXS map clearly shows the presence of a broad feature between $\sim$ 0.5 eV and 1.25 eV not observed at the $L_2$ edge.
This is seen more easily and in more detail when viewing the line scan cuts at resonance ($\mathrm{E_i}=11.961$ keV and 10.533 keV), shown in Fig. \ref{fig:fig9}.
Moving on now to focus on the broad $L_3$ edge feature, we note that the energy range of the feature roughly matches that of the spin-orbit excitations in \BYRO{}, which leads us to believe it shares the same origin, though broadened due to the itinerant nature of \SCRO{} at $\mathrm{T=300}$ K.
Besides the matching energy scale, the suppression of the feature at the $L_2$ edge mirrors that in \BYRO{} and agrees with its interpretation as spin-orbit coupling excitations involving the $L_2$-forbidden single-particle \jes{} level.
This interpretation of the broad $L_3$ feature as spin-orbit excitations also agrees with Marcaud et al.'s recently published $L_3$ edge RIXS measurements of \SCRO{} \cite{SCRO_RIXS}.
What's more, Marcaud et al.'s measurements of \SCRO{} reveal two peaks at $\sim$ 0.65 eV and 1 eV which are ascribed to Hund's coupled spin-orbit excitations, placing \SCRO{} in the intermediate Hund's coupling regime along with \BYRO{}.
Upon closer inspection, the broad SOC feature in our measurements of unstrained thin film \SCRO{} is slightly peaked around these energies, suggesting the feature may possess the same underlying doublet structure.

In order to better analyze the SOC feature, we fitted and subtracted off the $e-h$ continuum from the overall spectrum, shown in the zoomed inset of Fig. \ref{fig:fig9}.
Because the $dd$ excitations are suppressed at the $L_2$ edge we used this dataset to fit the $e-h$ continuum using an empirical decaying power law $f(x)=(x-x_0)^b+c$, yielding $b\sim -2.65,\ x_0\sim-1$ eV.
These parameters were then kept fixed when fitting the $L_3$ edge $e-h$ continuum, with only a constant background term and an overall scale term fitted.
The continuum subtracted results more clearly show that the $L_3$ edge spectrum appears to possess two SOC peaks at $\sim$ 0.65 eV and 1 eV, and furthermore reveal a third, weaker peak around 1.8 eV not reported by Marcaud et al..
Considering the low intensity of this third peak and its similar energy scale, we associate this feature with a `two-particle' excitation to the $^1S_0$ level as in \BYRO{}.
As described in our discussion of \BYRO{}, this third feature naturally appears in the intermediate Hund's coupling regime, confirming this same description is accurate for \SCRO{}.
Moreover, whereas the raw $L_2$ edge RIXS spectrum shows no apparent trace of the SOC features, the continuum subtracted spectrum appears to show a small bump in the region of the SOC features, suggesting that the excitations do not entirely disappear but are instead highly suppressed.
As discussed in $\S$\ref{sec:BYRO edge dep}, the fact that the SOC excitations do not disappear at the $L_2$ edge is again owed to the presence of Hund's coupling, which relaxes the strict single-particle $L_2$ edge selection rule.
While we can confidently state that \SCRO{} possesses strong SOC and intermediate Hund's coupling, the breadth of the features and weak $L_2$ signature makes it impossible to untangle their individual scales and discern whether \SCRO{} should possess $J_H/\zeta<1$ as reported by Marcaud et al. or whether it should also possess $J_H/\zeta>1$ as we have shown for \BYRO{} \cite{SCRO_RIXS}.

\section{\label{sec:discussion}Discussion}
Moving on from the preceding atomic theory analyses, we now discuss the possible origins of the features unaccounted for within this theory, beginning with the low energy feature in \BYRO{}.
As discussed in $\S$\ref{sec:BYRO common}, the resonance profile of this feature indicates that it either corresponds to a low energy $dd$ excitation or to a low energy non-$dd$ excitation.
The first case implies a splitting of the \Jtwo{} ground state, which agrees with theoretical works that propose a splitting $\delta$ of the ground state due to either a) the aforementioned $t_{2g}-e_g$ mixing effects or b) non-spherical Coulomb interactions allowed within an octahedral local environment  \cite{5d2_DPs_voleti,5d2_DPs_Paramekanti,5d2_DPs_Maharaj}.
Moreover, this splitting scheme theoretically helps to stabilize the recently reported charge-quadrupole and magnetic-dipole ordered states in \BYRO{} \cite{BYRO_order,BYRO_quadrupolar_order}.
On the other hand, $\delta$ is not anticipated to be large enough to account for the extent of the feature up to 250 meV in Fig. \ref{fig:fig3}.
Some estimates put $\delta$ at up to $\sim 50$ meV, while our RIXS calculations incorporating only the $t_{2g}-e_g$ mixing aspect show a splitting of only $\sim 15$ meV, see the Supplemental Material \cite{5d2_DPs_voleti,BYRO_quadrupolar_order,supp}.

Moving on to the second case of a low energy non-$dd$ excitation, one notable possibility due to visual similarity is that the feature represents ground state Jahn-Teller active phonon excitations (vibronic excitations) as seen in the $5d^1$ $\mathrm{A_2MgReO_6\ (A=Ca,\ Sr,\ Ba)}$ DPs \cite{Frontini_AMRO_2024}.
Indeed, $5d^2$ DPs such as \BYRO{} are theorized to also be subject to this so-called dynamic Jahn-Teller effect, and their RIXS spectra are expected to display a series of low energy excitations fitting the appearance of this feature \cite{5d2_vibronic_iwahara}.
Beyond a theoretical possibility, recent O $K$ edge RIXS measurements of the Os-based $5d^2$ DP $\mathrm{Ba_2CaOsO_6}$ clearly show a low energy phonon sideband fitting a dynamic Jahn-Teller description \cite{5d2_vibronic_iwahara,Ba2CaOsO6_5d2_vibronic}.
Moreover, the feature could potentially be explained as a combination of the prior two options, i.e. the feature corresponds to a $dd$ excitation across $\delta$ dressed by vibronic modes akin to the dressed SOC excitations seen in the aforementioned $\mathrm{A_2MgReO_6\ (A=Ca,\ Sr,\ Ba)}$ DPs.
In this case, the dressed excitation would extend to a significantly higher energy and could easily reconcile the theorized energy scale of $\delta$ with that of the feature observed here.

\begin{figure}[h]
\includegraphics[width=\columnwidth]{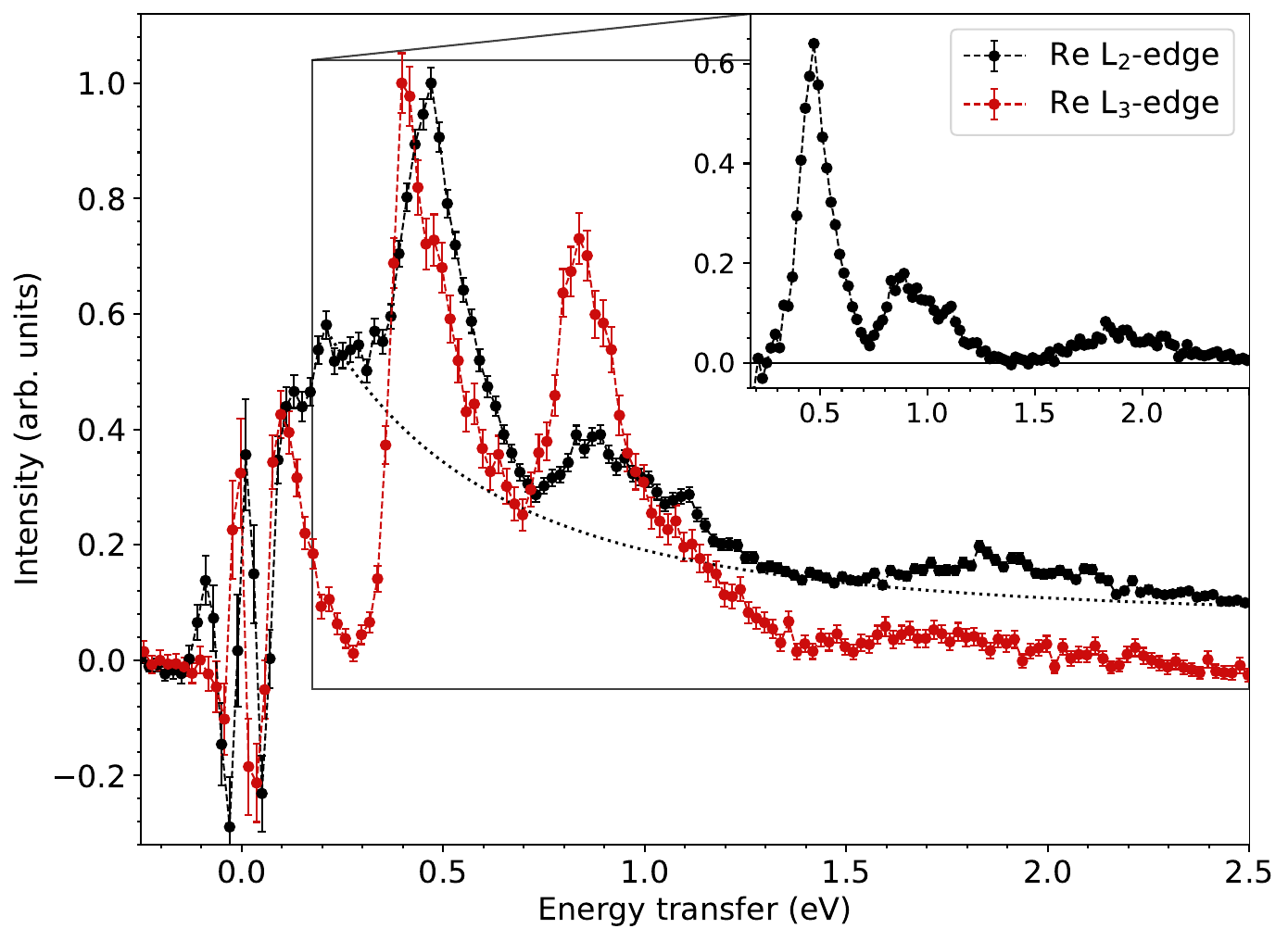}
\caption{\BYRO{} RIXS spectra at the Re $L_2$($L_3$) edge resonance of $\mathrm{E_i}=11.961$(10.533) keV up to 2.5 eV $\mathrm{E_t}$ with the elastic line subtracted. The data are overlaid with an $L_2$ edge decaying power law fit of the continuum background. The zoomed inset shows the $L_2$ spectrum with the continuum background subtracted. \label{fig:fig10}}
\end{figure}

In order to further elucidate the origin of the feature we scrutinize the edge-dependence of the feature shown in Fig. \ref{fig:fig4}, which reveals that the low energy region is significantly enhanced at the $L_2$ edge and appears to possess a different lineshape than at the $L_3$ edge.
We first note that this enhancement is not an illusory result of our normalization scheme and persists, though less dramatic, when the data are instead normalized by the elastic line, see the Supplemental Material \cite{supp}.
As for the apparent difference in lineshape between the two edges, we perform an elastic background subtraction to gain further insight, shown in Fig. \ref{fig:fig10}.
While the background subtracted spectra at both edges show a rise in inelastic intensity of a similar scale that begins around 100 meV, the feature at the $L_2$ edge extends continuously to the $^3P_1$ feature and does not decay unlike the feature at the $L_3$ edge.
The lineshape at the $L_2$ edge is moreover reminiscent of the continuum in \SCRO{}, and a fit to the same form of decaying power law appears to isolate the atomic features well, shown in the upper right inset of Fig. \ref{fig:fig10}.
The exact origin of this continuum is unclear; it may reflect an $e-h$ continuum as in \SCRO{}, or else could represent multi phonon processes as in a dynamic Jahn-Teller effect.
The former explanation is somewhat at odds with the supposedly insulating character of \BYRO{} according to first principle calculations \cite{BYRO_DFT_Kukusta}; however, we note that there is no experimental report of insulating behavior for \BYRO{}.
We also note that the insulating nature of \BYRO{} may be manifest in the fact that the continuum is much less prominent than in semiconducting \SCRO{}.
Unfortunately, both of these explanations also appear to be consistent with the observed edge dependence, making it difficult to draw any decisive conclusions.
In the case of an $e-h$ continuum, we observed similar edge dependence in \SCRO{} where the continuum was much stronger at the $L_2$ edge due to the partial suppression of the atomic features.
In the case of multi phonon excitations, recently presented theory results have also shown an enhancement of low energy vibronic excitations at the $L_2$ edge compared to the $L_3$ edge \cite{Naoya_5d1_L2}.
Due to the lack of clarity as to the nature of the continuum, it is also unclear whether the continuum and the low energy feature at the $L_3$ edge share the same origin, raising the possibility that there are two separate contributions to the low-energy lineshape.

Beside it's potential applicability to the low energy features, the presence of a dynamic Jahn-Teller effect may also explain some of the as-of-yet unaddressed quirks of our spectra that deviate from the theory to date.
In particular, both the 0.5 eV and 0.9 eV features appear distinctly asymmetric with additional spectral weight in the high energy region of the features reminiscent of the SOC excitations in the $\mathrm{A_2MgReO_6\ (A=Ca,\ Sr,\ Ba)}$ $5d^1$ DPs well described by vibronic coupling \cite{Frontini_AMRO_2024}.
Cementing this notion, vibronic theory for $5d^2$ DPs indeed predicts that such phonon sidebands should dress both the 0.5 eV and 0.9 eV features \cite{5d2_vibronic_iwahara}.
Additionally, we observe that the 0.9 eV feature is not only enhanced between $L_2$ and $L_3$ edges, but also changes in lineshape.
While the enhancement is understood, the change in lineshape does not mesh well with its explanation as a single $dd$ transition.
Conversely, if the high energy region instead corresponds to a phonon sideband then the change in lineshape can be explained by edge dependence of the vibronic excitations.
Thus, to summarize, the RIXS spectra of \BYRO{} and their edge dependence are well described by an atomic multiplet theory with $\zeta=0.290(5)$ eV and $J_H/\zeta=1.30(5)$ and by the presence of a dynamic Jahn-Teller effect possibly supplemented by an $e-h$ continuum.

Now shifting our focus to \SCRO{} and its deviation from the atomic theory, we expand on our preliminary discussion of the $e-h$ continuum with the additional detail provided by the line scans in Fig. \ref{fig:fig9}.
This additional detail immediately reveals the presence of a low energy shoulder between the elastic region and $\sim 200-300$ meV which does not follow the otherwise smooth trend of the $e-h$ continuum.
Notably, this feature is mirrored in Re $L_2$ edge RIXS measurements of the related $3d-5d$ high-$\mathrm{T_c}$ ferrimagnets \BFRO{} and \CFRO{}, leading us to believe that the feature may simply indicated to the beginning of the $e-h$ continuum \cite{BYRO_RIXS_Yuan2017}. 
Moreover, in comparison to the spectra of \BFRO{} and \CFRO{}, we note that the $L_2$ edge continuum is far more similar to that in metallic \BFRO{} than that in the larger band gap semiconductor \CFRO{} \cite{CFRO_insulating_BFRO_metal1,CFRO_insulating_BFRO_metal2}.
In particular, the shoulder feature in \BFRO{} merges with the elastic line as in \SCRO{}, whereas in \CFRO{} the shoulder feature becomes a true peak separated from the elastic.
Thus, if shoulder feature indeed represents the beginning of the $e-h$ continuum, merging vs. separation of the feature from the elastic should correspondingly reflect a gapless vs. gapped (or small gap vs. large gap) $e-h$ continuum.
This fits naturally with the respective transport properties of \BFRO{} and \CFRO{}, and should additionally reflect a near-metallic nature in  \SCRO{}.
Qualitatively, this is supported by the fact that \SCRO{} is much more conductive than the already semiconducting \CFRO{} \cite{SCRO_growth_Hauser2012,CFRO_insulating_BFRO_metal1,CFRO_insulating_BFRO_metal2}.
More quantitatively, this is also in line with the resistivity-estimated activation gap of 9.4 meV found in 200 nm thick \SCRO{} films, considering both that a temperature of T = 300 K would be sufficient to thermally populate the conduction band ($\mathrm{k_BT}\sim 26$ meV) and that this small of a gap would likely not be visible even at T = 0 K given our resolution of 65/95 meV FWHM.
On the other hand, our results agree less well with optical estimates of a 0.2 eV band gap which should be visible at room temperature and with our resolution \cite{SCRO_growth_Hauser2012}.
This latter point may indicate an indirect nature of the band gap given that optical probes typically only probe the direct gap due to the negligible momentum they can impart.
Conversely, hard X-rays can impart significant momentum and therefore probe an indirect band gap.

While we believe that the shoulder feature in \SCRO{} indeed corresponds to the beginning of the $e-h$ continuum, we now address other possibilities.
The first of these, which the reader may already have pondered, is that feature may be explained as multi-phonon excitations to match the low energy feature in \BYRO{}.
Indeed, we are initially encouraged by this possibility considering the energy scale and overall appearance of the feature is fairly similar to that of the feature in \BYRO{}.
Looking deeper, however, this option loses some plausibility as we do not observe any $L_2$ edge enhancement of the feature as we did in \BYRO{}, nor is it possible to look for hints in the lineshape of the SOC excitations given their broadness and overall lack of detail.
Thus, while we cannot reject this possibility with certainty, we do not think there is sufficient evidence to consider this the most likely origin of the feature.

Next, we discuss Marcaud et al.'s attribution of the feature to magnon excitations \cite{SCRO_RIXS}.
Specifically, Marcaud et al.'s calculations predict a zone boundary energy of a Re(Cr) magnon mode extending up to $\sim200(300)$ meV.
However, this energy scale far outstrips that of zone-boundary magnons measured by inelastic neutron scattering in the related high-$\mathrm{T_c}$ ferrimagnets \BFRO{} and \CFRO{} which extend up to a maximum energy of 50 meV \cite{SCRO_RIXS,BFRO_magnon,CFRO_magnon}.
Adding to this, both \BFRO{} and \CFRO{} order ferrimagnetically at temperatures on the same order of magnitude as \SCRO{} ($\mathrm{T_c}\sim305$ K and 520 K respectively), such that we anticipate their exchange couplings to be similar \cite{SCRO_2004_3d5dordering,CFRO_insulating_BFRO_metal1,BFRO_magnon,CFRO_magnon}.
Finally, we note that \BFRO{} and \CFRO{} possess B site spin moments of $S=5/2$ rather than $S=3/2$ as in \SCRO{} and thus possess significantly larger ferrimagnetic moment than \SCRO{} \cite{SCRO_2004_3d5dordering}.
Thus, on both counts of the moment size and exchange coupling we believe that it is unlikely that \SCRO{} should possess single-magnon bands $4-8$ times more energetic than those in \BFRO{} and \CFRO{}.
Moreover, regardless of the true magnon energy scale in \SCRO{}, we emphasize that \BFRO{} and \CFRO{} share similar low energy RIXS features to \SCRO{} that are decidedly incompatible with the energy scales of single-magnons in these materials. 
Thus, given this comparison, we are compelled to conclude that the low energy shoulder in \SCRO{} does not represent single-magnon excitations.
Although in principle multi-magnon modes can appear in RIXS \cite{RIXS_review}, we do not consider this a plausible origin for the feature because we would then correspondingly expect to see separate signatures of single-magnon excitations in the RIXS spectra of \SCRO{}, \BFRO{}, and \CFRO{} in the form of greater spectral weight in the region below 200 meV. 
\section{\label{sec:conclusion}Conclusion}
In summary, we have presented Re $L_2$ and $L_3$ edge RIXS measurements of the Re-based DPs \BYRO{} and \SCRO{} and have extracted powerful information from the differences between the two edges using simple single-particle selection rules.
In Mott-insulating \BYRO{}, these differences allowed us to discern that the hierarchy of the lowest excited states is in fact the inverse of what has previously been theorized, reflecting a need to revisit the appropriate energy scales of the Hund's coupling and SOC.
We used exact diagonalization methods to estimate said new energy scales and found that the values of $J_H=0.38(2)$ eV and $\zeta=0.290(5)$ eV describe our data well, which in particular puts $\zeta$ more in line with that found in other Re-based Mott insulators.

Our RIXS measurements of \BYRO{} also for the first time reveals the presence of edge dependent resonant low energy features which we conclude most plausibly represent Jahn-Teller active excitations, which have been predicted to occur in \BYRO{} and which have been observed in other Re-based DPs, and which may also be supplemented by an $e-h$ continuum.
The presence of a dynamic Jahn-Teller effect in \BYRO{} also helps explain the observed asymmetry and edge dependent lineshape of the 0.5 eV and 0.9 eV features.
Future RIXS experiments with higher resolution and expanded temperature dependence will be invaluable in refining our understand of vibronic physics in \BYRO{}.

In the correlated small band gap semiconductor \SCRO{}, the difference between $L_2$ and $L_3$ edge RIXS measurements confirms that the broad features between 0.5 eV and 1.5 eV correspond to Hund's coupled spin-orbit excitations.
Additionally, we have newly resolved the presence of `two-particle' spin-orbit excitations anticipated within the intermediate Hund's coupling regime.
Unfortunately, the presence of a strong $e-h$ continuum masks and broadens the $dd$ excitations in \SCRO{}, such that the individual scales of $J_H$ and $\zeta$ cannot be discerned.
Our measurements also reveal a near-metallic character in \SCRO{} as no discernible band gap exists in the $e-h$ continuum excitations.
A comparison with other works suggests the presence of an indirect band gap smaller than our instrumental resolution, in good agreement with prior estimates of a 9.4 meV activation gap.
Finally, the RIXS spectra of \BYRO{} and \SCRO{} reveal similar energy scales of Hund's coupling and SOC despite their respective Mott-insulating and semiconducting natures, reflecting the crucial role of B site substitution in $5d^2$ DPs.
\begin{acknowledgments}
Work at the University of Toronto was supported by the Natural Sciences and Engineering Research Council (NSERC) of Canada through the Discovery Grant No. RGPIN-2019-06449, Canada Foundation for Innovation, and Ontario Research Fund.
Work performed at Brookhaven National Laboratory was supported by the U.S. Department of Energy (DOE), Division of Materials Science, under Contract No. DE-SC0012704.
This research used resources of the Advanced Photon Source, a U.S. Department of Energy (DOE) Office of Science user facility operated for the DOE Office of Science by Argonne National Laboratory under Contract No. DE-AC02-06CH11357.
We acknowledge support from the US National Science Foundation (NSF) Grant Number 2201516 under the Accelnet program of Office of International Science and Engineering (OISE).
\end{acknowledgments}
\bibliography{main}

\begin{thebibliography}{47}%
\makeatletter
\providecommand \@ifxundefined [1]{%
 \@ifx{#1\undefined}
}%
\providecommand \@ifnum [1]{%
 \ifnum #1\expandafter \@firstoftwo
 \else \expandafter \@secondoftwo
 \fi
}%
\providecommand \@ifx [1]{%
 \ifx #1\expandafter \@firstoftwo
 \else \expandafter \@secondoftwo
 \fi
}%
\providecommand \natexlab [1]{#1}%
\providecommand \enquote  [1]{``#1''}%
\providecommand \bibnamefont  [1]{#1}%
\providecommand \bibfnamefont [1]{#1}%
\providecommand \citenamefont [1]{#1}%
\providecommand \href@noop [0]{\@secondoftwo}%
\providecommand \href [0]{\begingroup \@sanitize@url \@href}%
\providecommand \@href[1]{\@@startlink{#1}\@@href}%
\providecommand \@@href[1]{\endgroup#1\@@endlink}%
\providecommand \@sanitize@url [0]{\catcode `\\12\catcode `\$12\catcode `\&12\catcode `\#12\catcode `\^12\catcode `\_12\catcode `\%12\relax}%
\providecommand \@@startlink[1]{}%
\providecommand \@@endlink[0]{}%
\providecommand \url  [0]{\begingroup\@sanitize@url \@url }%
\providecommand \@url [1]{\endgroup\@href {#1}{\urlprefix }}%
\providecommand \urlprefix  [0]{URL }%
\providecommand \Eprint [0]{\href }%
\providecommand \doibase [0]{https://doi.org/}%
\providecommand \selectlanguage [0]{\@gobble}%
\providecommand \bibinfo  [0]{\@secondoftwo}%
\providecommand \bibfield  [0]{\@secondoftwo}%
\providecommand \translation [1]{[#1]}%
\providecommand \BibitemOpen [0]{}%
\providecommand \bibitemStop [0]{}%
\providecommand \bibitemNoStop [0]{.\EOS\space}%
\providecommand \EOS [0]{\spacefactor3000\relax}%
\providecommand \BibitemShut  [1]{\csname bibitem#1\endcsname}%
\let\auto@bib@innerbib\@empty
%</preamble>
\bibitem [{\citenamefont {Imada}\ \emph {et~al.}(1998)\citenamefont {Imada}, \citenamefont {Fujimori},\ and\ \citenamefont {Tokura}}]{Correlation_effects1}%
  \BibitemOpen
  \bibfield  {author} {\bibinfo {author} {\bibfnamefont {M.}~\bibnamefont {Imada}}, \bibinfo {author} {\bibfnamefont {A.}~\bibnamefont {Fujimori}},\ and\ \bibinfo {author} {\bibfnamefont {Y.}~\bibnamefont {Tokura}},\ }\bibfield  {title} {\bibinfo {title} {Metal-insulator transitions},\ }\href {https://doi.org/10.1103/RevModPhys.70.1039} {\bibfield  {journal} {\bibinfo  {journal} {Rev. Mod. Phys.}\ }\textbf {\bibinfo {volume} {70}},\ \bibinfo {pages} {1039} (\bibinfo {year} {1998})}\BibitemShut {NoStop}%
\bibitem [{\citenamefont {Vasala}\ and\ \citenamefont {Karppinen}(2015)}]{DPs_review}%
  \BibitemOpen
  \bibfield  {author} {\bibinfo {author} {\bibfnamefont {S.}~\bibnamefont {Vasala}}\ and\ \bibinfo {author} {\bibfnamefont {M.}~\bibnamefont {Karppinen}},\ }\bibfield  {title} {\bibinfo {title} {A2b'b"o6 perovskites: A review},\ }\href {https://doi.org/https://doi.org/10.1016/j.progsolidstchem.2014.08.001} {\bibfield  {journal} {\bibinfo  {journal} {Progress in Solid State Chemistry}\ }\textbf {\bibinfo {volume} {43}},\ \bibinfo {pages} {1} (\bibinfo {year} {2015})}\BibitemShut {NoStop}%
\bibitem [{\citenamefont {Chen}\ \emph {et~al.}(2010)\citenamefont {Chen}, \citenamefont {Pereira},\ and\ \citenamefont {Balents}}]{5d1_DPs_Chen}%
  \BibitemOpen
  \bibfield  {author} {\bibinfo {author} {\bibfnamefont {G.}~\bibnamefont {Chen}}, \bibinfo {author} {\bibfnamefont {R.}~\bibnamefont {Pereira}},\ and\ \bibinfo {author} {\bibfnamefont {L.}~\bibnamefont {Balents}},\ }\bibfield  {title} {\bibinfo {title} {Exotic phases induced by strong spin-orbit coupling in ordered double perovskites},\ }\href {https://doi.org/10.1103/PhysRevB.82.174440} {\bibfield  {journal} {\bibinfo  {journal} {Phys. Rev. B}\ }\textbf {\bibinfo {volume} {82}},\ \bibinfo {pages} {174440} (\bibinfo {year} {2010})}\BibitemShut {NoStop}%
\bibitem [{\citenamefont {Chen}\ and\ \citenamefont {Balents}(2011)}]{5d2_DPs_Chen}%
  \BibitemOpen
  \bibfield  {author} {\bibinfo {author} {\bibfnamefont {G.}~\bibnamefont {Chen}}\ and\ \bibinfo {author} {\bibfnamefont {L.}~\bibnamefont {Balents}},\ }\bibfield  {title} {\bibinfo {title} {Spin-orbit coupling in ${d}^{2}$ ordered double perovskites},\ }\href {https://doi.org/10.1103/PhysRevB.84.094420} {\bibfield  {journal} {\bibinfo  {journal} {Phys. Rev. B}\ }\textbf {\bibinfo {volume} {84}},\ \bibinfo {pages} {094420} (\bibinfo {year} {2011})}\BibitemShut {NoStop}%
\bibitem [{\citenamefont {Kobayashi}\ \emph {et~al.}(1998)\citenamefont {Kobayashi}, \citenamefont {Kimura}, \citenamefont {Sawada}, \citenamefont {Terakura},\ and\ \citenamefont {Tokura}}]{spintronics1}%
  \BibitemOpen
  \bibfield  {author} {\bibinfo {author} {\bibfnamefont {K.-I.}\ \bibnamefont {Kobayashi}}, \bibinfo {author} {\bibfnamefont {T.}~\bibnamefont {Kimura}}, \bibinfo {author} {\bibfnamefont {H.}~\bibnamefont {Sawada}}, \bibinfo {author} {\bibfnamefont {K.}~\bibnamefont {Terakura}},\ and\ \bibinfo {author} {\bibfnamefont {Y.}~\bibnamefont {Tokura}},\ }\bibfield  {title} {\bibinfo {title} {Room-temperature magnetoresistance in an oxide material with an ordered double-perovskite structure},\ }\href {https://doi.org/10.1038/27167} {\bibfield  {journal} {\bibinfo  {journal} {Nature}\ }\textbf {\bibinfo {volume} {395}},\ \bibinfo {pages} {677–680} (\bibinfo {year} {1998})}\BibitemShut {NoStop}%
\bibitem [{\citenamefont {\ifmmode \check{Z}\else \v{Z}\fi{}uti\ifmmode~\acute{c}\else \'{c}\fi{}}\ \emph {et~al.}(2004)\citenamefont {\ifmmode \check{Z}\else \v{Z}\fi{}uti\ifmmode~\acute{c}\else \'{c}\fi{}}, \citenamefont {Fabian},\ and\ \citenamefont {Das~Sarma}}]{spintronics2}%
  \BibitemOpen
  \bibfield  {author} {\bibinfo {author} {\bibfnamefont {I.}~\bibnamefont {\ifmmode \check{Z}\else \v{Z}\fi{}uti\ifmmode~\acute{c}\else \'{c}\fi{}}}, \bibinfo {author} {\bibfnamefont {J.}~\bibnamefont {Fabian}},\ and\ \bibinfo {author} {\bibfnamefont {S.}~\bibnamefont {Das~Sarma}},\ }\bibfield  {title} {\bibinfo {title} {Spintronics: Fundamentals and applications},\ }\href {https://doi.org/10.1103/RevModPhys.76.323} {\bibfield  {journal} {\bibinfo  {journal} {Rev. Mod. Phys.}\ }\textbf {\bibinfo {volume} {76}},\ \bibinfo {pages} {323} (\bibinfo {year} {2004})}\BibitemShut {NoStop}%
\bibitem [{\citenamefont {Serrate}\ \emph {et~al.}(2006)\citenamefont {Serrate}, \citenamefont {Teresa},\ and\ \citenamefont {Ibarra}}]{spintronics3}%
  \BibitemOpen
  \bibfield  {author} {\bibinfo {author} {\bibfnamefont {D.}~\bibnamefont {Serrate}}, \bibinfo {author} {\bibfnamefont {J.~M.~D.}\ \bibnamefont {Teresa}},\ and\ \bibinfo {author} {\bibfnamefont {M.~R.}\ \bibnamefont {Ibarra}},\ }\bibfield  {title} {\bibinfo {title} {Double perovskites with ferromagnetism above room temperature},\ }\href {https://doi.org/10.1088/0953-8984/19/2/023201} {\bibfield  {journal} {\bibinfo  {journal} {Journal of Physics: Condensed Matter}\ }\textbf {\bibinfo {volume} {19}},\ \bibinfo {pages} {023201} (\bibinfo {year} {2006})}\BibitemShut {NoStop}%
\bibitem [{\citenamefont {Paramekanti}\ \emph {et~al.}(2020)\citenamefont {Paramekanti}, \citenamefont {Maharaj},\ and\ \citenamefont {Gaulin}}]{5d2_DPs_Paramekanti}%
  \BibitemOpen
  \bibfield  {author} {\bibinfo {author} {\bibfnamefont {A.}~\bibnamefont {Paramekanti}}, \bibinfo {author} {\bibfnamefont {D.~D.}\ \bibnamefont {Maharaj}},\ and\ \bibinfo {author} {\bibfnamefont {B.~D.}\ \bibnamefont {Gaulin}},\ }\bibfield  {title} {\bibinfo {title} {Octupolar order in $d$-orbital mott insulators},\ }\href {https://doi.org/10.1103/PhysRevB.101.054439} {\bibfield  {journal} {\bibinfo  {journal} {Phys. Rev. B}\ }\textbf {\bibinfo {volume} {101}},\ \bibinfo {pages} {054439} (\bibinfo {year} {2020})}\BibitemShut {NoStop}%
\bibitem [{\citenamefont {Voleti}\ \emph {et~al.}(2020)\citenamefont {Voleti}, \citenamefont {Maharaj}, \citenamefont {Gaulin}, \citenamefont {Luke},\ and\ \citenamefont {Paramekanti}}]{5d2_DPs_voleti}%
  \BibitemOpen
  \bibfield  {author} {\bibinfo {author} {\bibfnamefont {S.}~\bibnamefont {Voleti}}, \bibinfo {author} {\bibfnamefont {D.~D.}\ \bibnamefont {Maharaj}}, \bibinfo {author} {\bibfnamefont {B.~D.}\ \bibnamefont {Gaulin}}, \bibinfo {author} {\bibfnamefont {G.}~\bibnamefont {Luke}},\ and\ \bibinfo {author} {\bibfnamefont {A.}~\bibnamefont {Paramekanti}},\ }\bibfield  {title} {\bibinfo {title} {Multipolar magnetism in $d$-orbital systems: Crystal field levels, octupolar order, and orbital loop currents},\ }\href {https://doi.org/10.1103/PhysRevB.101.155118} {\bibfield  {journal} {\bibinfo  {journal} {Phys. Rev. B}\ }\textbf {\bibinfo {volume} {101}},\ \bibinfo {pages} {155118} (\bibinfo {year} {2020})}\BibitemShut {NoStop}%
\bibitem [{\citenamefont {Witczak-Krempa}\ \emph {et~al.}(2014)\citenamefont {Witczak-Krempa}, \citenamefont {Chen}, \citenamefont {Kim},\ and\ \citenamefont {Balents}}]{Correlation_effects2}%
  \BibitemOpen
  \bibfield  {author} {\bibinfo {author} {\bibfnamefont {W.}~\bibnamefont {Witczak-Krempa}}, \bibinfo {author} {\bibfnamefont {G.}~\bibnamefont {Chen}}, \bibinfo {author} {\bibfnamefont {Y.~B.}\ \bibnamefont {Kim}},\ and\ \bibinfo {author} {\bibfnamefont {L.}~\bibnamefont {Balents}},\ }\bibfield  {title} {\bibinfo {title} {Correlated quantum phenomena in the strong spin-orbit regime},\ }\href {https://doi.org/https://doi.org/10.1146/annurev-conmatphys-020911-125138} {\bibfield  {journal} {\bibinfo  {journal} {Annual Review of Condensed Matter Physics}\ }\textbf {\bibinfo {volume} {5}},\ \bibinfo {pages} {57} (\bibinfo {year} {2014})}\BibitemShut {NoStop}%
\bibitem [{\citenamefont {Stamokostas}\ and\ \citenamefont {Fiete}(2018)}]{soc_eff_t2g_eg_mixing}%
  \BibitemOpen
  \bibfield  {author} {\bibinfo {author} {\bibfnamefont {G.~L.}\ \bibnamefont {Stamokostas}}\ and\ \bibinfo {author} {\bibfnamefont {G.~A.}\ \bibnamefont {Fiete}},\ }\bibfield  {title} {\bibinfo {title} {Mixing of ${t}_{2g}\ensuremath{-}{e}_{g}$ orbitals in $4d$ and $5d$ transition metal oxides},\ }\href {https://doi.org/10.1103/PhysRevB.97.085150} {\bibfield  {journal} {\bibinfo  {journal} {Phys. Rev. B}\ }\textbf {\bibinfo {volume} {97}},\ \bibinfo {pages} {085150} (\bibinfo {year} {2018})}\BibitemShut {NoStop}%
\bibitem [{\citenamefont {Georges}\ \emph {et~al.}(2013)\citenamefont {Georges}, \citenamefont {Medici},\ and\ \citenamefont {Mravlje}}]{HundsvsHubbard_metalvsins_Georges}%
  \BibitemOpen
  \bibfield  {author} {\bibinfo {author} {\bibfnamefont {A.}~\bibnamefont {Georges}}, \bibinfo {author} {\bibfnamefont {L.~d.}\ \bibnamefont {Medici}},\ and\ \bibinfo {author} {\bibfnamefont {J.}~\bibnamefont {Mravlje}},\ }\bibfield  {title} {\bibinfo {title} {Strong correlations from hund’s coupling},\ }\href {https://doi.org/https://doi.org/10.1146/annurev-conmatphys-020911-125045} {\bibfield  {journal} {\bibinfo  {journal} {Annual Review of Condensed Matter Physics}\ }\textbf {\bibinfo {volume} {4}},\ \bibinfo {pages} {137} (\bibinfo {year} {2013})}\BibitemShut {NoStop}%
\bibitem [{\citenamefont {Yuan}\ \emph {et~al.}(2017)\citenamefont {Yuan}, \citenamefont {Clancy}, \citenamefont {Cook}, \citenamefont {Thompson}, \citenamefont {Greedan}, \citenamefont {Cao}, \citenamefont {Jeon}, \citenamefont {Noh}, \citenamefont {Upton}, \citenamefont {Casa}, \citenamefont {Gog}, \citenamefont {Paramekanti},\ and\ \citenamefont {Kim}}]{BYRO_RIXS_Yuan2017}%
  \BibitemOpen
  \bibfield  {author} {\bibinfo {author} {\bibfnamefont {B.}~\bibnamefont {Yuan}}, \bibinfo {author} {\bibfnamefont {J.~P.}\ \bibnamefont {Clancy}}, \bibinfo {author} {\bibfnamefont {A.~M.}\ \bibnamefont {Cook}}, \bibinfo {author} {\bibfnamefont {C.~M.}\ \bibnamefont {Thompson}}, \bibinfo {author} {\bibfnamefont {J.}~\bibnamefont {Greedan}}, \bibinfo {author} {\bibfnamefont {G.}~\bibnamefont {Cao}}, \bibinfo {author} {\bibfnamefont {B.~C.}\ \bibnamefont {Jeon}}, \bibinfo {author} {\bibfnamefont {T.~W.}\ \bibnamefont {Noh}}, \bibinfo {author} {\bibfnamefont {M.~H.}\ \bibnamefont {Upton}}, \bibinfo {author} {\bibfnamefont {D.}~\bibnamefont {Casa}}, \bibinfo {author} {\bibfnamefont {T.}~\bibnamefont {Gog}}, \bibinfo {author} {\bibfnamefont {A.}~\bibnamefont {Paramekanti}},\ and\ \bibinfo {author} {\bibfnamefont {Y.-J.}\ \bibnamefont {Kim}},\ }\bibfield  {title} {\bibinfo {title} {Determination of hund's coupling in $5d$ oxides using resonant inelastic x-ray scattering},\ }\href
  {https://doi.org/10.1103/PhysRevB.95.235114} {\bibfield  {journal} {\bibinfo  {journal} {Phys. Rev. B}\ }\textbf {\bibinfo {volume} {95}},\ \bibinfo {pages} {235114} (\bibinfo {year} {2017})}\BibitemShut {NoStop}%
\bibitem [{\citenamefont {Nilsen}\ \emph {et~al.}(2021)\citenamefont {Nilsen}, \citenamefont {Thompson}, \citenamefont {Marjerisson}, \citenamefont {Badrtdinov}, \citenamefont {Tsirlin},\ and\ \citenamefont {Greedan}}]{BYRO_order}%
  \BibitemOpen
  \bibfield  {author} {\bibinfo {author} {\bibfnamefont {G.~J.}\ \bibnamefont {Nilsen}}, \bibinfo {author} {\bibfnamefont {C.~M.}\ \bibnamefont {Thompson}}, \bibinfo {author} {\bibfnamefont {C.}~\bibnamefont {Marjerisson}}, \bibinfo {author} {\bibfnamefont {D.~I.}\ \bibnamefont {Badrtdinov}}, \bibinfo {author} {\bibfnamefont {A.~A.}\ \bibnamefont {Tsirlin}},\ and\ \bibinfo {author} {\bibfnamefont {J.~E.}\ \bibnamefont {Greedan}},\ }\bibfield  {title} {\bibinfo {title} {Magnetic order and multipoles in the $5{d}^{2}$ rhenium double perovskite ${\mathrm{ba}}_{2}\mathrm{Y}\mathrm{Re}{\mathrm{o}}_{6}$},\ }\href {https://doi.org/10.1103/PhysRevB.103.104430} {\bibfield  {journal} {\bibinfo  {journal} {Phys. Rev. B}\ }\textbf {\bibinfo {volume} {103}},\ \bibinfo {pages} {104430} (\bibinfo {year} {2021})}\BibitemShut {NoStop}%
\bibitem [{\citenamefont {Omar}\ \emph {et~al.}(2025)\citenamefont {Omar}, \citenamefont {Zhang}, \citenamefont {Zhang}, \citenamefont {Tian}, \citenamefont {Dagotto}, \citenamefont {Chen}, \citenamefont {Arima}, \citenamefont {Stone}, \citenamefont {Christianson}, \citenamefont {Hirai},\ and\ \citenamefont {Gao}}]{BYRO_quadrupolar_order}%
  \BibitemOpen
  \bibfield  {author} {\bibinfo {author} {\bibfnamefont {O.}~\bibnamefont {Omar}}, \bibinfo {author} {\bibfnamefont {Y.}~\bibnamefont {Zhang}}, \bibinfo {author} {\bibfnamefont {Q.}~\bibnamefont {Zhang}}, \bibinfo {author} {\bibfnamefont {W.}~\bibnamefont {Tian}}, \bibinfo {author} {\bibfnamefont {E.}~\bibnamefont {Dagotto}}, \bibinfo {author} {\bibfnamefont {G.}~\bibnamefont {Chen}}, \bibinfo {author} {\bibfnamefont {T.-h.}\ \bibnamefont {Arima}}, \bibinfo {author} {\bibfnamefont {M.~B.}\ \bibnamefont {Stone}}, \bibinfo {author} {\bibfnamefont {A.~D.}\ \bibnamefont {Christianson}}, \bibinfo {author} {\bibfnamefont {D.}~\bibnamefont {Hirai}},\ and\ \bibinfo {author} {\bibfnamefont {S.}~\bibnamefont {Gao}},\ }\bibfield  {title} {\bibinfo {title} {Dipolar and quadrupolar correlations in the $5{d}^{2}$ re-based double perovskites ${\mathrm{ba}}_{2}{\mathrm{yreo}}_{6}$ and ${\mathrm{ba}}_{2}{\mathrm{screo}}_{6}$},\ }\href {https://doi.org/10.1103/gwph-w9zm} {\bibfield  {journal} {\bibinfo  {journal} {Phys. Rev.
  B}\ }\textbf {\bibinfo {volume} {112}},\ \bibinfo {pages} {075103} (\bibinfo {year} {2025})}\BibitemShut {NoStop}%
\bibitem [{\citenamefont {Aharen}\ \emph {et~al.}(2010)\citenamefont {Aharen}, \citenamefont {Greedan}, \citenamefont {Bridges}, \citenamefont {Aczel}, \citenamefont {Rodriguez}, \citenamefont {MacDougall}, \citenamefont {Luke}, \citenamefont {Michaelis}, \citenamefont {Kroeker}, \citenamefont {Wiebe}, \citenamefont {Zhou},\ and\ \citenamefont {Cranswick}}]{BYRO_synthesis_Aharen2010}%
  \BibitemOpen
  \bibfield  {author} {\bibinfo {author} {\bibfnamefont {T.}~\bibnamefont {Aharen}}, \bibinfo {author} {\bibfnamefont {J.~E.}\ \bibnamefont {Greedan}}, \bibinfo {author} {\bibfnamefont {C.~A.}\ \bibnamefont {Bridges}}, \bibinfo {author} {\bibfnamefont {A.~A.}\ \bibnamefont {Aczel}}, \bibinfo {author} {\bibfnamefont {J.}~\bibnamefont {Rodriguez}}, \bibinfo {author} {\bibfnamefont {G.}~\bibnamefont {MacDougall}}, \bibinfo {author} {\bibfnamefont {G.~M.}\ \bibnamefont {Luke}}, \bibinfo {author} {\bibfnamefont {V.~K.}\ \bibnamefont {Michaelis}}, \bibinfo {author} {\bibfnamefont {S.}~\bibnamefont {Kroeker}}, \bibinfo {author} {\bibfnamefont {C.~R.}\ \bibnamefont {Wiebe}}, \bibinfo {author} {\bibfnamefont {H.}~\bibnamefont {Zhou}},\ and\ \bibinfo {author} {\bibfnamefont {L.~M.~D.}\ \bibnamefont {Cranswick}},\ }\bibfield  {title} {\bibinfo {title} {Structure and magnetic properties of the $s=1$ geometrically frustrated double perovskites ${\text{la}}_{2}{\text{lireo}}_{6}$ and ${\text{ba}}_{2}{\text{yreo}}_{6}$},\
  }\href {https://doi.org/10.1103/PhysRevB.81.064436} {\bibfield  {journal} {\bibinfo  {journal} {Phys. Rev. B}\ }\textbf {\bibinfo {volume} {81}},\ \bibinfo {pages} {064436} (\bibinfo {year} {2010})}\BibitemShut {NoStop}%
\bibitem [{\citenamefont {Paramekanti}\ \emph {et~al.}(2018)\citenamefont {Paramekanti}, \citenamefont {Singh}, \citenamefont {Yuan}, \citenamefont {Casa}, \citenamefont {Said}, \citenamefont {Kim},\ and\ \citenamefont {Christianson}}]{BYRO_RIXS_Paramekanti2018}%
  \BibitemOpen
  \bibfield  {author} {\bibinfo {author} {\bibfnamefont {A.}~\bibnamefont {Paramekanti}}, \bibinfo {author} {\bibfnamefont {D.~J.}\ \bibnamefont {Singh}}, \bibinfo {author} {\bibfnamefont {B.}~\bibnamefont {Yuan}}, \bibinfo {author} {\bibfnamefont {D.}~\bibnamefont {Casa}}, \bibinfo {author} {\bibfnamefont {A.}~\bibnamefont {Said}}, \bibinfo {author} {\bibfnamefont {Y.-J.}\ \bibnamefont {Kim}},\ and\ \bibinfo {author} {\bibfnamefont {A.~D.}\ \bibnamefont {Christianson}},\ }\bibfield  {title} {\bibinfo {title} {Spin-orbit coupled systems in the atomic limit: rhenates, osmates, iridates},\ }\href {https://doi.org/10.1103/PhysRevB.97.235119} {\bibfield  {journal} {\bibinfo  {journal} {Phys. Rev. B}\ }\textbf {\bibinfo {volume} {97}},\ \bibinfo {pages} {235119} (\bibinfo {year} {2018})}\BibitemShut {NoStop}%
\bibitem [{\citenamefont {Blamire}\ \emph {et~al.}(2009)\citenamefont {Blamire}, \citenamefont {MacManus-Driscoll}, \citenamefont {Mathur},\ and\ \citenamefont {Barber}}]{AS_disorder_effect}%
  \BibitemOpen
  \bibfield  {author} {\bibinfo {author} {\bibfnamefont {M.~G.}\ \bibnamefont {Blamire}}, \bibinfo {author} {\bibfnamefont {J.~L.}\ \bibnamefont {MacManus-Driscoll}}, \bibinfo {author} {\bibfnamefont {N.~D.}\ \bibnamefont {Mathur}},\ and\ \bibinfo {author} {\bibfnamefont {Z.~H.}\ \bibnamefont {Barber}},\ }\bibfield  {title} {\bibinfo {title} {The materials science of functional oxide thin films},\ }\href {https://doi.org/https://doi.org/10.1002/adma.200900947} {\bibfield  {journal} {\bibinfo  {journal} {Advanced Materials}\ }\textbf {\bibinfo {volume} {21}},\ \bibinfo {pages} {3827} (\bibinfo {year} {2009})}\BibitemShut {NoStop}%
\bibitem [{\citenamefont {Kato}\ \emph {et~al.}(2002)\citenamefont {Kato}, \citenamefont {Okuda}, \citenamefont {Okimoto}, \citenamefont {Tomioka}, \citenamefont {Takenoya}, \citenamefont {Ohkubo}, \citenamefont {Kawasaki},\ and\ \citenamefont {Tokura}}]{SCRO_metal_growth2002}%
  \BibitemOpen
  \bibfield  {author} {\bibinfo {author} {\bibfnamefont {H.}~\bibnamefont {Kato}}, \bibinfo {author} {\bibfnamefont {T.}~\bibnamefont {Okuda}}, \bibinfo {author} {\bibfnamefont {Y.}~\bibnamefont {Okimoto}}, \bibinfo {author} {\bibfnamefont {Y.}~\bibnamefont {Tomioka}}, \bibinfo {author} {\bibfnamefont {Y.}~\bibnamefont {Takenoya}}, \bibinfo {author} {\bibfnamefont {A.}~\bibnamefont {Ohkubo}}, \bibinfo {author} {\bibfnamefont {M.}~\bibnamefont {Kawasaki}},\ and\ \bibinfo {author} {\bibfnamefont {Y.}~\bibnamefont {Tokura}},\ }\bibfield  {title} {\bibinfo {title} {{Metallic ordered double-perovskite Sr2CrReO6 with maximal Curie temperature of 635 K}},\ }\href {https://doi.org/10.1063/1.1493646} {\bibfield  {journal} {\bibinfo  {journal} {Applied Physics Letters}\ }\textbf {\bibinfo {volume} {81}},\ \bibinfo {pages} {328} (\bibinfo {year} {2002})}\BibitemShut {NoStop}%
\bibitem [{\citenamefont {Hauser}\ \emph {et~al.}(2012)\citenamefont {Hauser}, \citenamefont {Soliz}, \citenamefont {Dixit}, \citenamefont {Williams}, \citenamefont {Susner}, \citenamefont {Peters}, \citenamefont {Mier}, \citenamefont {Gustafson}, \citenamefont {Sumption}, \citenamefont {Fraser}, \citenamefont {Woodward},\ and\ \citenamefont {Yang}}]{SCRO_growth_Hauser2012}%
  \BibitemOpen
  \bibfield  {author} {\bibinfo {author} {\bibfnamefont {A.~J.}\ \bibnamefont {Hauser}}, \bibinfo {author} {\bibfnamefont {J.~R.}\ \bibnamefont {Soliz}}, \bibinfo {author} {\bibfnamefont {M.}~\bibnamefont {Dixit}}, \bibinfo {author} {\bibfnamefont {R.~E.~A.}\ \bibnamefont {Williams}}, \bibinfo {author} {\bibfnamefont {M.~A.}\ \bibnamefont {Susner}}, \bibinfo {author} {\bibfnamefont {B.}~\bibnamefont {Peters}}, \bibinfo {author} {\bibfnamefont {L.~M.}\ \bibnamefont {Mier}}, \bibinfo {author} {\bibfnamefont {T.~L.}\ \bibnamefont {Gustafson}}, \bibinfo {author} {\bibfnamefont {M.~D.}\ \bibnamefont {Sumption}}, \bibinfo {author} {\bibfnamefont {H.~L.}\ \bibnamefont {Fraser}}, \bibinfo {author} {\bibfnamefont {P.~M.}\ \bibnamefont {Woodward}},\ and\ \bibinfo {author} {\bibfnamefont {F.~Y.}\ \bibnamefont {Yang}},\ }\bibfield  {title} {\bibinfo {title} {Fully ordered sr${}_{2}$crreo${}_{6}$ epitaxial films: A high-temperature ferrimagnetic semiconductor},\ }\href {https://doi.org/10.1103/PhysRevB.85.161201}
  {\bibfield  {journal} {\bibinfo  {journal} {Phys. Rev. B}\ }\textbf {\bibinfo {volume} {85}},\ \bibinfo {pages} {161201} (\bibinfo {year} {2012})}\BibitemShut {NoStop}%
\bibitem [{\citenamefont {Chakraverty}\ \emph {et~al.}(2010)\citenamefont {Chakraverty}, \citenamefont {Ohtomo},\ and\ \citenamefont {Kawasaki}}]{AS_disorder_effect_SCRO}%
  \BibitemOpen
  \bibfield  {author} {\bibinfo {author} {\bibfnamefont {S.}~\bibnamefont {Chakraverty}}, \bibinfo {author} {\bibfnamefont {A.}~\bibnamefont {Ohtomo}},\ and\ \bibinfo {author} {\bibfnamefont {M.}~\bibnamefont {Kawasaki}},\ }\bibfield  {title} {\bibinfo {title} {{Controlled B-site ordering in Sr2CrReO6 double perovskite films by using pulsed laser interval deposition}},\ }\href {https://doi.org/10.1063/1.3525578} {\bibfield  {journal} {\bibinfo  {journal} {Applied Physics Letters}\ }\textbf {\bibinfo {volume} {97}},\ \bibinfo {pages} {243107} (\bibinfo {year} {2010})}\BibitemShut {NoStop}%
\bibitem [{\citenamefont {Esser}\ \emph {et~al.}(2016)\citenamefont {Esser}, \citenamefont {Hauser}, \citenamefont {Williams}, \citenamefont {Allen}, \citenamefont {Woodward}, \citenamefont {Yang},\ and\ \citenamefont {McComb}}]{SCRO_APD2017}%
  \BibitemOpen
  \bibfield  {author} {\bibinfo {author} {\bibfnamefont {B.~D.}\ \bibnamefont {Esser}}, \bibinfo {author} {\bibfnamefont {A.~J.}\ \bibnamefont {Hauser}}, \bibinfo {author} {\bibfnamefont {R.~E.~A.}\ \bibnamefont {Williams}}, \bibinfo {author} {\bibfnamefont {L.~J.}\ \bibnamefont {Allen}}, \bibinfo {author} {\bibfnamefont {P.~M.}\ \bibnamefont {Woodward}}, \bibinfo {author} {\bibfnamefont {F.~Y.}\ \bibnamefont {Yang}},\ and\ \bibinfo {author} {\bibfnamefont {D.~W.}\ \bibnamefont {McComb}},\ }\bibfield  {title} {\bibinfo {title} {Quantitative stem imaging of order-disorder phenomena in double perovskite thin films},\ }\href {https://doi.org/10.1103/PhysRevLett.117.176101} {\bibfield  {journal} {\bibinfo  {journal} {Phys. Rev. Lett.}\ }\textbf {\bibinfo {volume} {117}},\ \bibinfo {pages} {176101} (\bibinfo {year} {2016})}\BibitemShut {NoStop}%
\bibitem [{\citenamefont {Yuan}\ \emph {et~al.}(2021)\citenamefont {Yuan}, \citenamefont {Kim}, \citenamefont {Chun}, \citenamefont {Jin}, \citenamefont {Nelson}, \citenamefont {Hauser}, \citenamefont {Yang},\ and\ \citenamefont {Kim}}]{SCRO_moment_direction_Yuan2021}%
  \BibitemOpen
  \bibfield  {author} {\bibinfo {author} {\bibfnamefont {B.}~\bibnamefont {Yuan}}, \bibinfo {author} {\bibfnamefont {S.}~\bibnamefont {Kim}}, \bibinfo {author} {\bibfnamefont {S.~H.}\ \bibnamefont {Chun}}, \bibinfo {author} {\bibfnamefont {W.}~\bibnamefont {Jin}}, \bibinfo {author} {\bibfnamefont {C.~S.}\ \bibnamefont {Nelson}}, \bibinfo {author} {\bibfnamefont {A.~J.}\ \bibnamefont {Hauser}}, \bibinfo {author} {\bibfnamefont {F.~Y.}\ \bibnamefont {Yang}},\ and\ \bibinfo {author} {\bibfnamefont {Y.-J.}\ \bibnamefont {Kim}},\ }\bibfield  {title} {\bibinfo {title} {Robust long-range magnetic correlation across antiphase domain boundaries in ${\mathrm{sr}}_{2}{\mathrm{crreo}}_{6}$},\ }\href {https://doi.org/10.1103/PhysRevB.103.064410} {\bibfield  {journal} {\bibinfo  {journal} {Phys. Rev. B}\ }\textbf {\bibinfo {volume} {103}},\ \bibinfo {pages} {064410} (\bibinfo {year} {2021})}\BibitemShut {NoStop}%
\bibitem [{\citenamefont {Kato}\ \emph {et~al.}(2004)\citenamefont {Kato}, \citenamefont {Okuda}, \citenamefont {Okimoto}, \citenamefont {Tomioka}, \citenamefont {Oikawa}, \citenamefont {Kamiyama},\ and\ \citenamefont {Tokura}}]{SCRO_2004_3d5dordering}%
  \BibitemOpen
  \bibfield  {author} {\bibinfo {author} {\bibfnamefont {H.}~\bibnamefont {Kato}}, \bibinfo {author} {\bibfnamefont {T.}~\bibnamefont {Okuda}}, \bibinfo {author} {\bibfnamefont {Y.}~\bibnamefont {Okimoto}}, \bibinfo {author} {\bibfnamefont {Y.}~\bibnamefont {Tomioka}}, \bibinfo {author} {\bibfnamefont {K.}~\bibnamefont {Oikawa}}, \bibinfo {author} {\bibfnamefont {T.}~\bibnamefont {Kamiyama}},\ and\ \bibinfo {author} {\bibfnamefont {Y.}~\bibnamefont {Tokura}},\ }\bibfield  {title} {\bibinfo {title} {Structural and electronic properties of the ordered double perovskites ${A}_{2}m{\mathrm{reo}}_{6}$ $(a$=sr,ca; $m$=mg,sc,cr,mn,fe,co,ni,zn)},\ }\href {https://doi.org/10.1103/PhysRevB.69.184412} {\bibfield  {journal} {\bibinfo  {journal} {Phys. Rev. B}\ }\textbf {\bibinfo {volume} {69}},\ \bibinfo {pages} {184412} (\bibinfo {year} {2004})}\BibitemShut {NoStop}%
\bibitem [{\citenamefont {Vaitheeswaran}\ \emph {et~al.}(2005)\citenamefont {Vaitheeswaran}, \citenamefont {Kanchana},\ and\ \citenamefont {Delin}}]{SCRO_pseudo_half_metal_2005}%
  \BibitemOpen
  \bibfield  {author} {\bibinfo {author} {\bibfnamefont {G.}~\bibnamefont {Vaitheeswaran}}, \bibinfo {author} {\bibfnamefont {V.}~\bibnamefont {Kanchana}},\ and\ \bibinfo {author} {\bibfnamefont {A.}~\bibnamefont {Delin}},\ }\bibfield  {title} {\bibinfo {title} {Pseudo-half-metallicity in the double perovskite sr2crreo6 from density-functional calculations},\ }\href {https://doi.org/10.1063/1.1855418} {\bibfield  {journal} {\bibinfo  {journal} {Applied Physics Letters}\ }\textbf {\bibinfo {volume} {86}},\ \bibinfo {pages} {032513} (\bibinfo {year} {2005})}\BibitemShut {NoStop}%
\bibitem [{\citenamefont {Meetei}\ \emph {et~al.}(2013)\citenamefont {Meetei}, \citenamefont {Erten}, \citenamefont {Randeria}, \citenamefont {Trivedi},\ and\ \citenamefont {Woodward}}]{HundsvsHubbard_3d35d3_Meetei}%
  \BibitemOpen
  \bibfield  {author} {\bibinfo {author} {\bibfnamefont {O.~N.}\ \bibnamefont {Meetei}}, \bibinfo {author} {\bibfnamefont {O.}~\bibnamefont {Erten}}, \bibinfo {author} {\bibfnamefont {M.}~\bibnamefont {Randeria}}, \bibinfo {author} {\bibfnamefont {N.}~\bibnamefont {Trivedi}},\ and\ \bibinfo {author} {\bibfnamefont {P.}~\bibnamefont {Woodward}},\ }\bibfield  {title} {\bibinfo {title} {Theory of high ${T}_{c}$ ferrimagnetism in a multiorbital mott insulator},\ }\href {https://doi.org/10.1103/PhysRevLett.110.087203} {\bibfield  {journal} {\bibinfo  {journal} {Phys. Rev. Lett.}\ }\textbf {\bibinfo {volume} {110}},\ \bibinfo {pages} {087203} (\bibinfo {year} {2013})}\BibitemShut {NoStop}%
\bibitem [{\citenamefont {Chen}(2020)}]{HundsvsSOC_metalvsins_gangchen}%
  \BibitemOpen
  \bibfield  {author} {\bibinfo {author} {\bibfnamefont {G.}~\bibnamefont {Chen}},\ }\bibfield  {title} {\bibinfo {title} {Dilemma in strongly correlated materials: Hund's metal vs relativistic mott insulator},\ }\Eprint {https://arxiv.org/abs/2012.06752} {arXiv:2012.06752 [cond-mat.str-el]}  (\bibinfo {year} {2020})\BibitemShut {NoStop}%
\bibitem [{\citenamefont {Marcaud}\ \emph {et~al.}(2023)\citenamefont {Marcaud}, \citenamefont {Taekyung~Lee}, \citenamefont {Hauser}, \citenamefont {Yang}, \citenamefont {Lee}, \citenamefont {Casa}, \citenamefont {Upton}, \citenamefont {Gog}, \citenamefont {Saritas}, \citenamefont {Wang}, \citenamefont {Dean}, \citenamefont {Zhou}, \citenamefont {Zhang}, \citenamefont {Walker}, \citenamefont {Jarrige}, \citenamefont {Ismail-Beigi},\ and\ \citenamefont {Ahn}}]{SCRO_RIXS}%
  \BibitemOpen
  \bibfield  {author} {\bibinfo {author} {\bibfnamefont {G.}~\bibnamefont {Marcaud}}, \bibinfo {author} {\bibfnamefont {A.}~\bibnamefont {Taekyung~Lee}}, \bibinfo {author} {\bibfnamefont {A.~J.}\ \bibnamefont {Hauser}}, \bibinfo {author} {\bibfnamefont {F.~Y.}\ \bibnamefont {Yang}}, \bibinfo {author} {\bibfnamefont {S.}~\bibnamefont {Lee}}, \bibinfo {author} {\bibfnamefont {D.}~\bibnamefont {Casa}}, \bibinfo {author} {\bibfnamefont {M.}~\bibnamefont {Upton}}, \bibinfo {author} {\bibfnamefont {T.}~\bibnamefont {Gog}}, \bibinfo {author} {\bibfnamefont {K.}~\bibnamefont {Saritas}}, \bibinfo {author} {\bibfnamefont {Y.}~\bibnamefont {Wang}}, \bibinfo {author} {\bibfnamefont {M.~P.~M.}\ \bibnamefont {Dean}}, \bibinfo {author} {\bibfnamefont {H.}~\bibnamefont {Zhou}}, \bibinfo {author} {\bibfnamefont {Z.}~\bibnamefont {Zhang}}, \bibinfo {author} {\bibfnamefont {F.~J.}\ \bibnamefont {Walker}}, \bibinfo {author} {\bibfnamefont {I.}~\bibnamefont {Jarrige}}, \bibinfo {author} {\bibfnamefont {S.}~\bibnamefont
  {Ismail-Beigi}},\ and\ \bibinfo {author} {\bibfnamefont {C.}~\bibnamefont {Ahn}},\ }\bibfield  {title} {\bibinfo {title} {Low-energy electronic interactions in ferrimagnetic ${\mathrm{sr}}_{2}\mathrm{CrRe}{\mathrm{o}}_{6}$ thin films},\ }\href {https://doi.org/10.1103/PhysRevB.108.075132} {\bibfield  {journal} {\bibinfo  {journal} {Phys. Rev. B}\ }\textbf {\bibinfo {volume} {108}},\ \bibinfo {pages} {075132} (\bibinfo {year} {2023})}\BibitemShut {NoStop}%
\bibitem [{\citenamefont {Frontini}\ \emph {et~al.}(2024)\citenamefont {Frontini}, \citenamefont {Johnstone}, \citenamefont {Iwahara}, \citenamefont {Bhattacharyya}, \citenamefont {Bogdanov}, \citenamefont {Hozoi}, \citenamefont {Upton}, \citenamefont {Casa}, \citenamefont {Hirai},\ and\ \citenamefont {Kim}}]{Frontini_AMRO_2024}%
  \BibitemOpen
  \bibfield  {author} {\bibinfo {author} {\bibfnamefont {F.~I.}\ \bibnamefont {Frontini}}, \bibinfo {author} {\bibfnamefont {G.~H.~J.}\ \bibnamefont {Johnstone}}, \bibinfo {author} {\bibfnamefont {N.}~\bibnamefont {Iwahara}}, \bibinfo {author} {\bibfnamefont {P.}~\bibnamefont {Bhattacharyya}}, \bibinfo {author} {\bibfnamefont {N.~A.}\ \bibnamefont {Bogdanov}}, \bibinfo {author} {\bibfnamefont {L.}~\bibnamefont {Hozoi}}, \bibinfo {author} {\bibfnamefont {M.~H.}\ \bibnamefont {Upton}}, \bibinfo {author} {\bibfnamefont {D.~M.}\ \bibnamefont {Casa}}, \bibinfo {author} {\bibfnamefont {D.}~\bibnamefont {Hirai}},\ and\ \bibinfo {author} {\bibfnamefont {Y.-J.}\ \bibnamefont {Kim}},\ }\bibfield  {title} {\bibinfo {title} {Spin-orbit-lattice entangled state in ${\mathrm{a}}_{2}{\mathrm{mgreo}}_{6}$ ($\mathrm{A}=\mathrm{Ca}$, sr, ba) revealed by resonant inelastic x-ray scattering},\ }\href {https://doi.org/10.1103/PhysRevLett.133.036501} {\bibfield  {journal} {\bibinfo  {journal} {Phys. Rev. Lett.}\ }\textbf {\bibinfo
  {volume} {133}},\ \bibinfo {pages} {036501} (\bibinfo {year} {2024})}\BibitemShut {NoStop}%
\bibitem [{\citenamefont {Iwahara}\ \emph {et~al.}(2025)\citenamefont {Iwahara}, \citenamefont {Soh}, \citenamefont {Hirai}, \citenamefont {Živković}, \citenamefont {Wei}, \citenamefont {Zhang}, \citenamefont {Galdino}, \citenamefont {Yu}, \citenamefont {Ishii}, \citenamefont {Pisani}, \citenamefont {Malanyuk}, \citenamefont {Schmitt},\ and\ \citenamefont {Rønnow}}]{Ba2CaReO6_5d1_vibronic}%
  \BibitemOpen
  \bibfield  {author} {\bibinfo {author} {\bibfnamefont {N.}~\bibnamefont {Iwahara}}, \bibinfo {author} {\bibfnamefont {J.-R.}\ \bibnamefont {Soh}}, \bibinfo {author} {\bibfnamefont {D.}~\bibnamefont {Hirai}}, \bibinfo {author} {\bibfnamefont {I.}~\bibnamefont {Živković}}, \bibinfo {author} {\bibfnamefont {Y.}~\bibnamefont {Wei}}, \bibinfo {author} {\bibfnamefont {W.}~\bibnamefont {Zhang}}, \bibinfo {author} {\bibfnamefont {C.}~\bibnamefont {Galdino}}, \bibinfo {author} {\bibfnamefont {T.}~\bibnamefont {Yu}}, \bibinfo {author} {\bibfnamefont {K.}~\bibnamefont {Ishii}}, \bibinfo {author} {\bibfnamefont {F.}~\bibnamefont {Pisani}}, \bibinfo {author} {\bibfnamefont {O.}~\bibnamefont {Malanyuk}}, \bibinfo {author} {\bibfnamefont {T.}~\bibnamefont {Schmitt}},\ and\ \bibinfo {author} {\bibfnamefont {H.~M.}\ \bibnamefont {Rønnow}},\ }\bibfield  {title} {\bibinfo {title} {Persistent quantum vibronic dynamics in a $5d^1$ double perovskite oxide},\ }\href {https://doi.org/10.1103/vjtk-jsdg} {\bibfield  {journal}
  {\bibinfo  {journal} {Phys. Rev. B}\ ,\ \bibinfo {pages} {in press}} (\bibinfo {year} {2025})}\BibitemShut {NoStop}%
\bibitem [{\citenamefont {Warzanowski}\ \emph {et~al.}(2024)\citenamefont {Warzanowski}, \citenamefont {Magnaterra}, \citenamefont {Schlicht}, \citenamefont {Faure}, \citenamefont {Sahle}, \citenamefont {Becker}, \citenamefont {Bohat\'y}, \citenamefont {Sala}, \citenamefont {Monaco}, \citenamefont {Hermanns}, \citenamefont {van Loosdrecht},\ and\ \citenamefont {Gr\"uninger}}]{5d3_K2ReCl6_SOC}%
  \BibitemOpen
  \bibfield  {author} {\bibinfo {author} {\bibfnamefont {P.}~\bibnamefont {Warzanowski}}, \bibinfo {author} {\bibfnamefont {M.}~\bibnamefont {Magnaterra}}, \bibinfo {author} {\bibfnamefont {G.}~\bibnamefont {Schlicht}}, \bibinfo {author} {\bibfnamefont {Q.}~\bibnamefont {Faure}}, \bibinfo {author} {\bibfnamefont {C.~J.}\ \bibnamefont {Sahle}}, \bibinfo {author} {\bibfnamefont {P.}~\bibnamefont {Becker}}, \bibinfo {author} {\bibfnamefont {L.}~\bibnamefont {Bohat\'y}}, \bibinfo {author} {\bibfnamefont {M.~M.}\ \bibnamefont {Sala}}, \bibinfo {author} {\bibfnamefont {G.}~\bibnamefont {Monaco}}, \bibinfo {author} {\bibfnamefont {M.}~\bibnamefont {Hermanns}}, \bibinfo {author} {\bibfnamefont {P.~H.~M.}\ \bibnamefont {van Loosdrecht}},\ and\ \bibinfo {author} {\bibfnamefont {M.}~\bibnamefont {Gr\"uninger}},\ }\bibfield  {title} {\bibinfo {title} {Spin-orbit coupling in a half-filled ${t}_{2g}$ shell: The case of $5{d}^{3}$ ${\mathrm{k}}_{2}{\mathrm{recl}}_{6}$},\ }\href {https://doi.org/10.1103/PhysRevB.109.155149}
  {\bibfield  {journal} {\bibinfo  {journal} {Phys. Rev. B}\ }\textbf {\bibinfo {volume} {109}},\ \bibinfo {pages} {155149} (\bibinfo {year} {2024})}\BibitemShut {NoStop}%
\bibitem [{\citenamefont {Iwahara}(2024{\natexlab{a}})}]{5d2_vibronic_iwahara}%
  \BibitemOpen
  \bibfield  {author} {\bibinfo {author} {\bibfnamefont {N.}~\bibnamefont {Iwahara}},\ }\bibfield  {title} {\bibinfo {title} {Dynamic jahn–teller phenomena in heavy transition metal compounds},\ }\href {https://doi.org/10.7566/JPSJ.93.121003} {\bibfield  {journal} {\bibinfo  {journal} {Journal of the Physical Society of Japan}\ }\textbf {\bibinfo {volume} {93}},\ \bibinfo {pages} {121003} (\bibinfo {year} {2024}{\natexlab{a}})}\BibitemShut {NoStop}%
\bibitem [{sup()}]{supp}%
  \BibitemOpen
  \href@noop {} {}\bibinfo {note} {See Supplemental Material at [URL provided by publisher] for additional data and details of theoretical calculations}\BibitemShut {NoStop}%
\bibitem [{\citenamefont {Cowan}(1981)}]{Cowan_Theory_1981}%
  \BibitemOpen
  \bibfield  {author} {\bibinfo {author} {\bibfnamefont {R.~D.}\ \bibnamefont {Cowan}},\ }\href@noop {} {\emph {\bibinfo {title} {The {Theory} of {Atomic} {Structure} and {Spectra}}}},\ \bibinfo {series} {Los {Alamos} {Series} in {Basic} and {Applied} {Sciences}}, Vol.~\bibinfo {volume} {3}\ (\bibinfo  {publisher} {University of California Press},\ \bibinfo {year} {1981})\BibitemShut {NoStop}%
\bibitem [{\citenamefont {König}(1971)}]{konig_nephelauxetic_1971}%
  \BibitemOpen
  \bibfield  {author} {\bibinfo {author} {\bibfnamefont {E.}~\bibnamefont {König}},\ }\bibfield  {title} {\bibinfo {title} {The nephelauxetic effect calculation and accuracy of the interelectronic repulsion parameters {I}. {Cubic} high-spin d2, d3, d7, and d8 systems},\ }in\ \href {https://doi.org/10.1007/BFb0118887} {\emph {\bibinfo {booktitle} {Structutal and {Bonding}}}}\ (\bibinfo  {publisher} {Springer},\ \bibinfo {address} {Berlin, Heidelberg},\ \bibinfo {year} {1971})\ pp.\ \bibinfo {pages} {175--212}\BibitemShut {NoStop}%
\bibitem [{\citenamefont {Ament}\ \emph {et~al.}(2011)\citenamefont {Ament}, \citenamefont {Khaliullin},\ and\ \citenamefont {van~den Brink}}]{L2_selection_rules_iridium_Ament}%
  \BibitemOpen
  \bibfield  {author} {\bibinfo {author} {\bibfnamefont {L.~J.~P.}\ \bibnamefont {Ament}}, \bibinfo {author} {\bibfnamefont {G.}~\bibnamefont {Khaliullin}},\ and\ \bibinfo {author} {\bibfnamefont {J.}~\bibnamefont {van~den Brink}},\ }\bibfield  {title} {\bibinfo {title} {Theory of resonant inelastic x-ray scattering in iridium oxide compounds: Probing spin-orbit-entangled ground states and excitations},\ }\href {https://doi.org/10.1103/PhysRevB.84.020403} {\bibfield  {journal} {\bibinfo  {journal} {Phys. Rev. B}\ }\textbf {\bibinfo {volume} {84}},\ \bibinfo {pages} {020403} (\bibinfo {year} {2011})}\BibitemShut {NoStop}%
\bibitem [{\citenamefont {Clancy}\ \emph {et~al.}(2012)\citenamefont {Clancy}, \citenamefont {Chen}, \citenamefont {Kim}, \citenamefont {Chen}, \citenamefont {Plumb}, \citenamefont {Jeon}, \citenamefont {Noh},\ and\ \citenamefont {Kim}}]{L2_selection_rules_iridium_Clancy}%
  \BibitemOpen
  \bibfield  {author} {\bibinfo {author} {\bibfnamefont {J.~P.}\ \bibnamefont {Clancy}}, \bibinfo {author} {\bibfnamefont {N.}~\bibnamefont {Chen}}, \bibinfo {author} {\bibfnamefont {C.~Y.}\ \bibnamefont {Kim}}, \bibinfo {author} {\bibfnamefont {W.~F.}\ \bibnamefont {Chen}}, \bibinfo {author} {\bibfnamefont {K.~W.}\ \bibnamefont {Plumb}}, \bibinfo {author} {\bibfnamefont {B.~C.}\ \bibnamefont {Jeon}}, \bibinfo {author} {\bibfnamefont {T.~W.}\ \bibnamefont {Noh}},\ and\ \bibinfo {author} {\bibfnamefont {Y.-J.}\ \bibnamefont {Kim}},\ }\bibfield  {title} {\bibinfo {title} {Spin-orbit coupling in iridium-based 5$d$ compounds probed by x-ray absorption spectroscopy},\ }\href {https://doi.org/10.1103/PhysRevB.86.195131} {\bibfield  {journal} {\bibinfo  {journal} {Phys. Rev. B}\ }\textbf {\bibinfo {volume} {86}},\ \bibinfo {pages} {195131} (\bibinfo {year} {2012})}\BibitemShut {NoStop}%
\bibitem [{\citenamefont {Wang}\ \emph {et~al.}(2019)\citenamefont {Wang}, \citenamefont {Fabbris}, \citenamefont {Dean},\ and\ \citenamefont {Kotliar}}]{EDRIXS_WANG2019}%
  \BibitemOpen
  \bibfield  {author} {\bibinfo {author} {\bibfnamefont {Y.}~\bibnamefont {Wang}}, \bibinfo {author} {\bibfnamefont {G.}~\bibnamefont {Fabbris}}, \bibinfo {author} {\bibfnamefont {M.}~\bibnamefont {Dean}},\ and\ \bibinfo {author} {\bibfnamefont {G.}~\bibnamefont {Kotliar}},\ }\bibfield  {title} {\bibinfo {title} {Edrixs: An open source toolkit for simulating spectra of resonant inelastic x-ray scattering},\ }\href {https://doi.org/https://doi.org/10.1016/j.cpc.2019.04.018} {\bibfield  {journal} {\bibinfo  {journal} {Computer Physics Communications}\ }\textbf {\bibinfo {volume} {243}},\ \bibinfo {pages} {151} (\bibinfo {year} {2019})}\BibitemShut {NoStop}%
\bibitem [{\citenamefont {Maharaj}\ \emph {et~al.}(2020)\citenamefont {Maharaj}, \citenamefont {Sala}, \citenamefont {Stone}, \citenamefont {Kermarrec}, \citenamefont {Ritter}, \citenamefont {Fauth}, \citenamefont {Marjerrison}, \citenamefont {Greedan}, \citenamefont {Paramekanti},\ and\ \citenamefont {Gaulin}}]{5d2_DPs_Maharaj}%
  \BibitemOpen
  \bibfield  {author} {\bibinfo {author} {\bibfnamefont {D.~D.}\ \bibnamefont {Maharaj}}, \bibinfo {author} {\bibfnamefont {G.}~\bibnamefont {Sala}}, \bibinfo {author} {\bibfnamefont {M.~B.}\ \bibnamefont {Stone}}, \bibinfo {author} {\bibfnamefont {E.}~\bibnamefont {Kermarrec}}, \bibinfo {author} {\bibfnamefont {C.}~\bibnamefont {Ritter}}, \bibinfo {author} {\bibfnamefont {F.}~\bibnamefont {Fauth}}, \bibinfo {author} {\bibfnamefont {C.~A.}\ \bibnamefont {Marjerrison}}, \bibinfo {author} {\bibfnamefont {J.~E.}\ \bibnamefont {Greedan}}, \bibinfo {author} {\bibfnamefont {A.}~\bibnamefont {Paramekanti}},\ and\ \bibinfo {author} {\bibfnamefont {B.~D.}\ \bibnamefont {Gaulin}},\ }\bibfield  {title} {\bibinfo {title} {Octupolar versus n\'eel order in cubic $5{d}^{2}$ double perovskites},\ }\href {https://doi.org/10.1103/PhysRevLett.124.087206} {\bibfield  {journal} {\bibinfo  {journal} {Phys. Rev. Lett.}\ }\textbf {\bibinfo {volume} {124}},\ \bibinfo {pages} {087206} (\bibinfo {year} {2020})}\BibitemShut {NoStop}%
\bibitem [{\citenamefont {Okamoto}\ \emph {et~al.}(2025)\citenamefont {Okamoto}, \citenamefont {Shibata}, \citenamefont {Ponosov}, \citenamefont {Hayashi}, \citenamefont {Yamaura}, \citenamefont {Huang}, \citenamefont {Singh}, \citenamefont {Chen}, \citenamefont {Tanaka}, \citenamefont {Streltsov}, \citenamefont {Huang},\ and\ \citenamefont {Fujimori}}]{Ba2CaOsO6_5d2_vibronic}%
  \BibitemOpen
  \bibfield  {author} {\bibinfo {author} {\bibfnamefont {J.}~\bibnamefont {Okamoto}}, \bibinfo {author} {\bibfnamefont {G.}~\bibnamefont {Shibata}}, \bibinfo {author} {\bibfnamefont {Y.~S.}\ \bibnamefont {Ponosov}}, \bibinfo {author} {\bibfnamefont {H.}~\bibnamefont {Hayashi}}, \bibinfo {author} {\bibfnamefont {K.}~\bibnamefont {Yamaura}}, \bibinfo {author} {\bibfnamefont {H.~Y.}\ \bibnamefont {Huang}}, \bibinfo {author} {\bibfnamefont {A.}~\bibnamefont {Singh}}, \bibinfo {author} {\bibfnamefont {C.~T.}\ \bibnamefont {Chen}}, \bibinfo {author} {\bibfnamefont {A.}~\bibnamefont {Tanaka}}, \bibinfo {author} {\bibfnamefont {S.~V.}\ \bibnamefont {Streltsov}}, \bibinfo {author} {\bibfnamefont {D.~J.}\ \bibnamefont {Huang}},\ and\ \bibinfo {author} {\bibfnamefont {A.}~\bibnamefont {Fujimori}},\ }\bibfield  {title} {\bibinfo {title} {Spin-orbit-entangled electronic structure of ba$_2$caoso$_6$ studied by o $k$-edge resonant inelastic x-ray scattering and raman spectroscopy},\ }\href
  {https://doi.org/10.1038/s41535-025-00757-4} {\bibfield  {journal} {\bibinfo  {journal} {npj Quantum Materials}\ }\textbf {\bibinfo {volume} {10}},\ \bibinfo {pages} {44} (\bibinfo {year} {2025})}\BibitemShut {NoStop}%
\bibitem [{\citenamefont {Kukusta}\ \emph {et~al.}(2025)\citenamefont {Kukusta}, \citenamefont {Bekenov},\ and\ \citenamefont {Antonov}}]{BYRO_DFT_Kukusta}%
  \BibitemOpen
  \bibfield  {author} {\bibinfo {author} {\bibfnamefont {D.}~\bibnamefont {Kukusta}}, \bibinfo {author} {\bibfnamefont {L.}~\bibnamefont {Bekenov}},\ and\ \bibinfo {author} {\bibfnamefont {V.}~\bibnamefont {Antonov}},\ }\bibfield  {title} {\bibinfo {title} {Resonant inelastic x-ray scattering in double perovskites from first-principles. i. a2yreo6 (a = ba and sr)},\ }\href {https://doi.org/https://doi.org/10.1016/j.jmmm.2024.172714} {\bibfield  {journal} {\bibinfo  {journal} {Journal of Magnetism and Magnetic Materials}\ }\textbf {\bibinfo {volume} {614}},\ \bibinfo {pages} {172714} (\bibinfo {year} {2025})}\BibitemShut {NoStop}%
\bibitem [{\citenamefont {Iwahara}(2024{\natexlab{b}})}]{Naoya_5d1_L2}%
  \BibitemOpen
  \bibfield  {author} {\bibinfo {author} {\bibfnamefont {N.}~\bibnamefont {Iwahara}},\ }\bibfield  {title} {\bibinfo {title} {Theory of dynamic jahn-teller effect in 5d1 double perovskites and rixs spectra}} (\bibinfo {year} {2024}{\natexlab{b}}),\ \bibinfo {note} {physical Society of Japan, Spring Meeting}\BibitemShut {NoStop}%
\bibitem [{\citenamefont {Prellier}\ \emph {et~al.}(2000)\citenamefont {Prellier}, \citenamefont {Smolyaninova}, \citenamefont {Biswas}, \citenamefont {Galley}, \citenamefont {Greene}, \citenamefont {Ramesha},\ and\ \citenamefont {Gopalakrishnan}}]{CFRO_insulating_BFRO_metal1}%
  \BibitemOpen
  \bibfield  {author} {\bibinfo {author} {\bibfnamefont {W.}~\bibnamefont {Prellier}}, \bibinfo {author} {\bibfnamefont {V.}~\bibnamefont {Smolyaninova}}, \bibinfo {author} {\bibfnamefont {A.}~\bibnamefont {Biswas}}, \bibinfo {author} {\bibfnamefont {C.}~\bibnamefont {Galley}}, \bibinfo {author} {\bibfnamefont {R.~L.}\ \bibnamefont {Greene}}, \bibinfo {author} {\bibfnamefont {K.}~\bibnamefont {Ramesha}},\ and\ \bibinfo {author} {\bibfnamefont {J.}~\bibnamefont {Gopalakrishnan}},\ }\bibfield  {title} {\bibinfo {title} {Properties of the ferrimagnetic double perovskites a2fereo6 (a = ba and ca)},\ }\href {https://doi.org/10.1088/0953-8984/12/6/318} {\bibfield  {journal} {\bibinfo  {journal} {Journal of Physics: Condensed Matter}\ }\textbf {\bibinfo {volume} {12}},\ \bibinfo {pages} {965} (\bibinfo {year} {2000})}\BibitemShut {NoStop}%
\bibitem [{\citenamefont {Jeon}\ \emph {et~al.}(2010)\citenamefont {Jeon}, \citenamefont {Kim}, \citenamefont {Moon}, \citenamefont {Choi}, \citenamefont {Jeong}, \citenamefont {Lee}, \citenamefont {Yu}, \citenamefont {Won}, \citenamefont {Jung}, \citenamefont {Hur},\ and\ \citenamefont {Noh}}]{CFRO_insulating_BFRO_metal2}%
  \BibitemOpen
  \bibfield  {author} {\bibinfo {author} {\bibfnamefont {B.~C.}\ \bibnamefont {Jeon}}, \bibinfo {author} {\bibfnamefont {C.~H.}\ \bibnamefont {Kim}}, \bibinfo {author} {\bibfnamefont {S.~J.}\ \bibnamefont {Moon}}, \bibinfo {author} {\bibfnamefont {W.~S.}\ \bibnamefont {Choi}}, \bibinfo {author} {\bibfnamefont {H.}~\bibnamefont {Jeong}}, \bibinfo {author} {\bibfnamefont {Y.~S.}\ \bibnamefont {Lee}}, \bibinfo {author} {\bibfnamefont {J.}~\bibnamefont {Yu}}, \bibinfo {author} {\bibfnamefont {C.~J.}\ \bibnamefont {Won}}, \bibinfo {author} {\bibfnamefont {J.~H.}\ \bibnamefont {Jung}}, \bibinfo {author} {\bibfnamefont {N.}~\bibnamefont {Hur}},\ and\ \bibinfo {author} {\bibfnamefont {T.~W.}\ \bibnamefont {Noh}},\ }\bibfield  {title} {\bibinfo {title} {Electronic structure of double perovskite a2fereo6 (a = ba and ca): interplay between spin–orbit interaction, electron correlation, and lattice distortion},\ }\href {https://doi.org/10.1088/0953-8984/22/34/345602} {\bibfield  {journal} {\bibinfo  {journal} {Journal
  of Physics: Condensed Matter}\ }\textbf {\bibinfo {volume} {22}},\ \bibinfo {pages} {345602} (\bibinfo {year} {2010})}\BibitemShut {NoStop}%
\bibitem [{\citenamefont {Plumb}\ \emph {et~al.}(2013)\citenamefont {Plumb}, \citenamefont {Cook}, \citenamefont {Clancy}, \citenamefont {Kolesnikov}, \citenamefont {Jeon}, \citenamefont {Noh}, \citenamefont {Paramekanti},\ and\ \citenamefont {Kim}}]{BFRO_magnon}%
  \BibitemOpen
  \bibfield  {author} {\bibinfo {author} {\bibfnamefont {K.~W.}\ \bibnamefont {Plumb}}, \bibinfo {author} {\bibfnamefont {A.~M.}\ \bibnamefont {Cook}}, \bibinfo {author} {\bibfnamefont {J.~P.}\ \bibnamefont {Clancy}}, \bibinfo {author} {\bibfnamefont {A.~I.}\ \bibnamefont {Kolesnikov}}, \bibinfo {author} {\bibfnamefont {B.~C.}\ \bibnamefont {Jeon}}, \bibinfo {author} {\bibfnamefont {T.~W.}\ \bibnamefont {Noh}}, \bibinfo {author} {\bibfnamefont {A.}~\bibnamefont {Paramekanti}},\ and\ \bibinfo {author} {\bibfnamefont {Y.-J.}\ \bibnamefont {Kim}},\ }\bibfield  {title} {\bibinfo {title} {Neutron scattering study of magnetic excitations in a 5$d$-based double-perovskite ba${}_{2}$fereo${}_{6}$},\ }\href {https://doi.org/10.1103/PhysRevB.87.184412} {\bibfield  {journal} {\bibinfo  {journal} {Phys. Rev. B}\ }\textbf {\bibinfo {volume} {87}},\ \bibinfo {pages} {184412} (\bibinfo {year} {2013})}\BibitemShut {NoStop}%
\bibitem [{\citenamefont {Yuan}\ \emph {et~al.}(2018)\citenamefont {Yuan}, \citenamefont {Clancy}, \citenamefont {Sears}, \citenamefont {Kolesnikov}, \citenamefont {Stone}, \citenamefont {Yamani}, \citenamefont {Won}, \citenamefont {Hur}, \citenamefont {Jeon}, \citenamefont {Noh}, \citenamefont {Paramekanti},\ and\ \citenamefont {Kim}}]{CFRO_magnon}%
  \BibitemOpen
  \bibfield  {author} {\bibinfo {author} {\bibfnamefont {B.}~\bibnamefont {Yuan}}, \bibinfo {author} {\bibfnamefont {J.~P.}\ \bibnamefont {Clancy}}, \bibinfo {author} {\bibfnamefont {J.~A.}\ \bibnamefont {Sears}}, \bibinfo {author} {\bibfnamefont {A.~I.}\ \bibnamefont {Kolesnikov}}, \bibinfo {author} {\bibfnamefont {M.~B.}\ \bibnamefont {Stone}}, \bibinfo {author} {\bibfnamefont {Z.}~\bibnamefont {Yamani}}, \bibinfo {author} {\bibfnamefont {C.}~\bibnamefont {Won}}, \bibinfo {author} {\bibfnamefont {N.}~\bibnamefont {Hur}}, \bibinfo {author} {\bibfnamefont {B.~C.}\ \bibnamefont {Jeon}}, \bibinfo {author} {\bibfnamefont {T.~W.}\ \bibnamefont {Noh}}, \bibinfo {author} {\bibfnamefont {A.}~\bibnamefont {Paramekanti}},\ and\ \bibinfo {author} {\bibfnamefont {Y.-J.}\ \bibnamefont {Kim}},\ }\bibfield  {title} {\bibinfo {title} {Neutron scattering investigation of rhenium orbital ordering in the $3d\ensuremath{-}5d$ double perovskite ${\mathrm{ca}}_{2}{\mathrm{fereo}}_{6}$},\ }\href
  {https://doi.org/10.1103/PhysRevB.98.214433} {\bibfield  {journal} {\bibinfo  {journal} {Phys. Rev. B}\ }\textbf {\bibinfo {volume} {98}},\ \bibinfo {pages} {214433} (\bibinfo {year} {2018})}\BibitemShut {NoStop}%
\bibitem [{\citenamefont {Mitrano}\ \emph {et~al.}(2024)\citenamefont {Mitrano}, \citenamefont {Johnston}, \citenamefont {Kim},\ and\ \citenamefont {Dean}}]{RIXS_review}%
  \BibitemOpen
  \bibfield  {author} {\bibinfo {author} {\bibfnamefont {M.}~\bibnamefont {Mitrano}}, \bibinfo {author} {\bibfnamefont {S.}~\bibnamefont {Johnston}}, \bibinfo {author} {\bibfnamefont {Y.-J.}\ \bibnamefont {Kim}},\ and\ \bibinfo {author} {\bibfnamefont {M.~P.~M.}\ \bibnamefont {Dean}},\ }\bibfield  {title} {\bibinfo {title} {Exploring quantum materials with resonant inelastic x-ray scattering},\ }\href {https://doi.org/10.1103/PhysRevX.14.040501} {\bibfield  {journal} {\bibinfo  {journal} {Phys. Rev. X}\ }\textbf {\bibinfo {volume} {14}},\ \bibinfo {pages} {040501} (\bibinfo {year} {2024})}\BibitemShut {NoStop}%
\end{thebibliography}%

\end{document}